# Local structure of glassy lithium phosphorus oxynitride thin films: a combined experimental and *ab initio* approach

Maxwell A.T. Marple*[a], Thomas A. Wynn[b], Diyi Cheng[c], Ryosuke Shimizu[b], Harris E. Mason[a], and Y. Shirley Meng*[b],[c],[d]

[a] M.A.T. Marple, H. E. Mason
Physical and Life Science Directorate
Lawrence Livermore National Laboratory
Livermore, CA 94550, United States
E-mail: marple1@llnl.gov

[b] T.A. Wynn, R. Shimizu, Prof. Y. S. Meng
Department of NanoEngineering
University of California San Diego

[c] D. Cheng, Prof. Y. S. Meng
Materials Science & Engineering Program
University of California San Diego

[d] Prof. Y. S. Meng
Sustainable Power and Energy Center
University of California San Diego
La Jolla, CA 92093, United States
E-mail: shirleymeng@ucsd.edu

Supporting information for this article is given via a link at the end of the document.

**Abstract:** Lithium phosphorus oxynitride (LiPON) is an amorphous solid-state lithium ion conductor displaying exemplary cyclability against lithium metal anodes. There is no definitive explanation for this stability due to the limited understanding of the structure of LiPON. We provide a structural model of RF-sputtered LiPON via experimental and computational spectroscopic methods. Information about the short-range structure results from 1D and 2D solid-state nuclear magnetic resonance experiments investigating chemical shift anisotropy and dipolar interactions. These results are compared with first principles chemical shielding calculations of Li-P-O/N crystals and *ab initio* molecular dynamics-generated amorphous LiPON models to unequivocally identify the glassy structure as primarily isolated phosphate monomers with N incorporated in both apical and as bridging sites in phosphate dimers. Structural results suggest LiPON's stability is a result of its glassy character. Free-standing LiPON films are produced that exhibit a high degree of flexibility highlighting the unique mechanical properties of glassy materials.

## Introduction

Solid-state lithium ion conductors are attractive electrolytes for next generation lithium ion batteries due to their improved safety and their potential to improve energy density by enabling the use of lithium metal anodes.[1] However, in practice the use of Li metal anodes is hindered by electrolyte decomposition and the formation of high impedance interfaces[2] or the formation of Li dendrites.[3] This degradation diminishes coulombic efficiency and Li dendrite formation causes catastrophic failure via cell shorting. A fundamental understanding of solid electrolytes relative stability against Li metal is required to surmount the issue of degradation; however, the relevant properties leading to stability is currently disputed. Standard descriptions of the electrochemical stability window of solid-state interfaces rely on calculations of the grand potential phase diagrams to compute interface stability, drawing parallels to the solid electrolyte interphase in liquid electrolytes.[2,4] However, mechanisms of electrochemical decomposition are incomplete, and generally do not incorporate kinetics into these models, with recent exceptions.[5]

While many material properties of crystalline compounds can be predicted by thermodynamic calculations, the same is not true for glassy materials as they are non-equilibrium and their formation and properties are largely driven by kinetics. Traditionally a glass is formed after atomic motion is kinetically arrested during the rapid quench from a melt. The configurational state that is frozen is a local potential energy minimum within a potential energy landscape and the subsequent thermodynamic properties of the glass are dictated by the local potential energy minima and transitions between these minima govern the relaxation and transport kinetics.[6] The glass is metastable as the kinetics for crystallization or decomposition become impossibly slow on any reasonable, 'human' time scale. As kinetics are paramount to glass properties, by not considering kinetics in electrochemical decomposition models the response of glassy solid electrolytes will be incorrect. Notably, glassy solid electrolytes are of interest because they lack grain boundaries that can be sources of electrostatic and structural inhomogeneities[7] and charge transfer impedance,[8] and they can have wider compositional stability than analogous crystals so they may tolerate ionic depletion and not undergo a phase transformation.[9] However, glassy electrolytes such as lithium phosphorus oxynitride (LiPON) and lithium thiophoshates are both experimentally and computationally challenging to investigate: experimentally, methods for explicit structural determination are limited; computationally, disordered solids are more difficult given the prevalence of periodic boundary conditions in most theory. LiPON, the focus of this work, is particularly interesting due to its remarkable cyclability against lithium metal—a crucial requirement for next-generation lithium ion batteries.[10] LiPON's stability has been attributed to a number of features: low electronic conductivity ($10^{-15}$-$10^{-12}$ S cm$^{-1}$)[3], mechanical rigidity[11], formation of electrically insulating and ionically conducting decomposition products, $Li_3P$, $Li_3N$, $Li_2O$, supported by density functional theory (DFT) predictions and *in situ* x-ray photoelectron spectroscopy *(*XPS)[2,12,13], and kinetic stability of those interfacial components.[14]

Despite the exemplary electrochemical stability LiPON displays, there is a pervasive lack of understanding and inconsistency in describing the local structure of LiPON. Many of these comments are already discussed elsewhere with much of the confusion stemming from how N is incorporated into the structure and the types of local structural units.[15,16] These issues arise from observations made on metaphosphate oxynitride glasses where XPS results indicated N crosslinks the network by bonding to three and two P tetrahedra denoted $N_t$ and $N_d$, respectively. Others claim structural models of LiPON include extended chain structures where many P tetrahedra are linked by bridging O or N similar to metaphosphate glasses or even a layered structure of Li and P rich regions.[13,17–20] These descriptions are inconsistent with the structure of LiPON's precursor material $Li_3PO_4$ that is composed of isolated P tetrahedra. As a result, existing kinetic models for the Li/LiPON interfaces overestimate structural instability through an overabundance of such metastable coordination environments.[14,19] In this regard, accurate local structural descriptions are of utmost importance in describing the chemical environments leading to the enhanced stability. Recent investigations have shown through a combination of neutron scattering and *ab initio* molecular dynamics (AIMD) that N is incorporated into LiPON by forming both dimeric $P_2O_6N^{5-}$ units where N is bridging ($N_d$) and a non-bridging N site on isolated $PO_3N^{4-}$ units (apical N, $N_a$).[15,21] They find no evidence of $N_t$ and offer alternative assignments for XPS and IR spectroscopic results in support of their findings.

Solid-state nuclear magnetic resonance (NMR) spectroscopy is particularly suited for the determination of structure in glasses as it is sensitive to short range order and can probe a number of interactions like chemical shift anisotropy (CSA), dipolar and quadrupolar coupling that contain unique information to help distinguish different chemical environments within an amorphous material. Notably it offers quantitative insight into the constituent short range structural units of LiPON and can validate recently proposed structural models.[22] The typical connectivity nomenclature for phosphate glasses is given by the number of bridging oxygens (BO) per tetrahedral P atom, $Q^n$, where *n* is the number of bridging atoms and ranges from 3 to 0. A network composed of $Q^3$ units is three dimensional, whereas a $Q^2$ network is defined by chains, $Q^1$ network is composed solely of dimeric units, and $Q^0$ by isolated $PO_4^{3-}$ tetrahedra.[23] Modifying cations like Li act to depolymerize the network by forming non-bridging oxygen (NBO) atoms randomly throughout the network, thus direct insight into the network connectivity is gained by tracking the population of the Q units, and the $Q^n$ speciation corresponds directly to the Li:P ratio. To account for the mixed-anion effect on connectivity we introduce a modification of the $Q^n$ nomenclature, $Q^n_m$ where *m* is the number of non-oxygen anions on the P tetrahedra and can take on values between 0 to 4. In the case of LiPON, *m* indicates the number of nitrogen atoms per P tetrahedra as N substitutes O when it is incorporated into the glass network.[24] Other NMR investigations have been performed on LiPON, however these studies focused on LiPON synthesized by atypical deposition methods and on bulk LiPON glasses closer to metaphosphate compositions.[25,26] In this work, we employ advanced 2D NMR techniques to differentiate the local chemical shift anisotropic features and dipolar interactions permitting structural determination to resolve the local structure of RF-sputtered LiPON. The experimental NMR results are compared to those from density functional theory using the gauge included projector augmented wave (GIPAW) framework to calculate the chemical shieldings of a variety of lithium phosphorus oxynitride crystals, developing a database of local bonding environments and their corresponding chemical shielding. These calculated shielding values and CSA parameters are compared to the measured chemical shifts from NMR measurements, and are then used to validate AIMD-amorphized LiPON structure. This combined experimental and computational study provides unique information of the local structure and is consistent with recently proposed structural models, thus providing a definitive and unequivocal local structural model of LiPON.[15,21] Such structural validation is crucial for developing an atomic level understanding of the electrochemical stability of this electrolyte when paired with lithium metal.

## Results and Discussion

### $^{31}P$ MAS NMR of LiPON

NMR is sensitive to short range structure, making it an invaluable tool for structural identification of glasses wherein the connectivity of structural units is resolved as separate chemical shifts. For investigating structure, $^{31}P$ NMR is favorable as it is a sensitive nucleus to local chemical environments and is a 100% abundant. The deconvolution and subsequent interpretation of the $^{31}P$ spectra of LiPON is non-trivial and requires the simultaneous consideration of the experimental and calculated results to create a consistent structural model. The $^{31}P$ magic angle spinning (MAS) spectrum of LiPON (Fig. 1) shows a broad peak centered at 10 ppm with a high frequency shoulder. The line shape is broadened by structural disorder arising from a distribution of bond lengths and angles and consequently a chemical shift distribution.

To aid in interpreting the $^{31}P$ line shape of LiPON, spectra of the target material, crystalline β-$Li_3PO_4$ (c-LPO), and RF-sputtered amorphous films of $Li_3PO_4$ (a-LPO) are collected (Fig. S3). The $^{31}P$ chemical shift for c-LPO has a sharp peak at 9.6 ppm in accordance with orthophosphate tetrahedra having four non-bridging oxygen, $Q^0$. After amorphization, a-LPO shows a broadening of the $Q^0$ peak indicating structural disorder. A shoulder is observable at ~0 ppm that can be attributed to dimeric $P_2O_7$ units, $Q^1$, suggesting about 3 mol% $Li_2O$ is lost during the sputtering process, as alkali phosphate glasses follow a random binary distribution in terms of Q speciation with cation concentration.[27] The chemical shift for $Q^1$ is slightly higher than observed in a pyrophosphate crystal[28], likely a result of a redistribution of the electron density with higher cation concentration. Increases of up to 10 ppm for a Q$^n$ species with increasing cation concentration have been previously observed.[29] The MAS $^{31}$P spectrum of a LiPON film (Figure 1a) bears resemblance to the a-LPO spectra with the predominant intensity at 9.1 ppm and a small tail around 3 ppm; these sites can be comfortably assigned to Q$^0_0$ and Q$^1_0$ phosphate species, respectively. However, in contrast to the a-LPO film, there are additional shoulders to higher chemical shift that are presumed to be phosphorus bonded to N. Further assignments are hindered by a lack of understanding of how N is incorporated into alkali pyro- and orthophosphate glasses. Previous studies on the effect of nitridation in metaphosphate glasses have found a similar rise of higher frequency peaks that are attributed to forming various PO$_3$N and PO$_2$N$_2$ Q$^n_m$ units.[26,30,31]

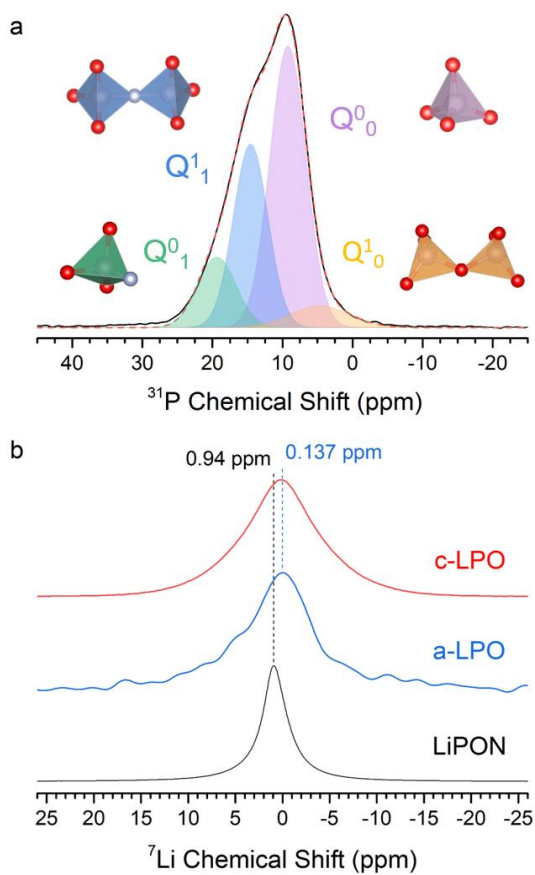

**Figure 1.** (a) $^{31}$P MAS NMR spectra of free-standing thin film LiPON spinning at 25 kHz, (b) $^{7}$Li MAS NMR spectra of powder Li$_3$PO$_4$, thin film amorphous Li$_3$PO$_4$, and free-standing thin film LiPON.

We turn to first principles calculations of a database of lithium phosphorus oxynitride compounds (Tables S4-S7) and AIMD simulations of LiPON (Table 1, SI) to aid our assignments. These methods are discussed in the following section and detailed in Supplementary Information (SI). Our results indicate the $^{31}$P spectra of LiPON (Fig. 1) can be deconvoluted into 4 peaks, the majority of which is composed of Q$^0_0$ PO$_4^{3-}$ units at 9.3 ppm, followed by Q$^1_1$ P$_2$O$_6$N$^{5-}$ dimer units in which N is bridging two PO$_3$N tetrahedra whose $\delta_{iso}$=14.6 ppm; the other nitride species at 19.4 ppm is assigned to Q$^0_1$ PO$_3$N$^{4-}$ units, and a minor amount of Q$^1_0$ P$_2$O$_7^{4-}$ dimers as previously mentioned. These assignments give important insight into how N is incorporated into the LPO network, suggesting that N acts similarly to O, as both a bridging (as observed in the Q$^1_1$ site) and non-bridging (the Q$^0_1$ site) anion. The overall line shape is similar to that of previously published $^{31}$P NMR of IBAD deposited LiPON, with similar peak positions and broadening that is associated with N incorporation.[25] However, the IBAD spectrum shows diminished intensity above 12 ppm in comparison, indicating the network has less Q$^1_1$ and no Q$^0_1$ units. The observation of entrapped N$_2$ gas within the IBAD film agrees with this finding that less N is incorporated with the IBAD process.

While $^{31}$P NMR offers insight into the network structure, $^{6}$Li (Figure S4) and $^{7}$Li (Figure 1b) chemical shifts of c/a-LPO and LiPON were collected to understand the changes of the Li environments. Both $^{6}$Li and $^{7}$Li chemical shifts of LiPON increase by ~1 ppm relative to c-LPO and a-LPO as a result of the lowering Li coordination to less than 4.[22] The $^{7}$Li line shape of LiPON is narrowed relative to a/c-LPO suggesting increased Li conductivity, in agreement with dielectric spectroscopy measurements.[32] It should be noted, the $^{7}$Li LiPON line shape has a Lorentzian character indicating the Li ions are mobile at room temperature and rapidly exchanging between sites, hence a single peak is observed.

### Computational spectroscopy of lithium oxynitride phosphates

To accurately correlate NMR chemical shifts with local structures and remove ambiguities in the assignments of chemical environments, we employ DFT calculations to simulate the effective electronic shielding of a variety of relevant lithium phosphorus oxynitride compounds. Recent implementation of the GIPAW approach has enabled precise determination of electronic shielding effects on nuclei in solids[33,34] that directly relate to chemical shifts determined via NMR measurements.[35] The chemical shielding is related to chemical shift by a correlation factor based on experimentally determined chemical shifts (SI). With this correlation factor, isotropic chemical shifts, $\delta_{iso}$, and CSA parameters can be calculated for all the structures in the database allowing for systematic trends of the structures with chemical shift to be observed. The VASP implementation of the GIPAW approach was applied to a body of lithium phosphorus oxynitride crystals of varied compositions and bonding environments; the list of compounds and their calculated chemical shifts are shown in Tables S4-S7. To confirm prediction of known structural groups, calculations were performed for Li$_3$PO$_4$ (Q$^0$ PO$_4$), Li$_4$P$_2$O$_7$ (Q$^1$ PO$_4$ dimers), and LiPO$_3$ (Q$^2$ PO$_4$ chains), along with some phosphorus nitride and oxynitride variants, shown in Figure 2; corrected $\delta_{iso}$ (Figure S12) are accurately predicted in comparison to experimentally determined chemical shifts (Table S5). By organizing all the compounds in the database by their Q$^n$ speciation and anion type (Fig. 2b, left), there is a clear trend showing that as Q$^n$ is reduced the $^{31}$P chemical shift correspondingly increases by about 16-20 ppm. This agrees with previous observations of Q$^n$ speciation trends in phosphate glasses[36] and shows that the Q$^n$ speciation largely dictates $\delta_{iso}$. While the database lacks many representative data points, the effect of nitridation has a clear effect on $\delta_{iso}$ although weaker in comparison to the effect of Q$^n$. As the variation of $\delta_{iso}$ of the Q$^0_m$ units shows, $\delta_{iso}$ increases with increasing nitridation (Q$_m$) by about 10 ppm for every N replacing O, and likely applies to all Q$^n$

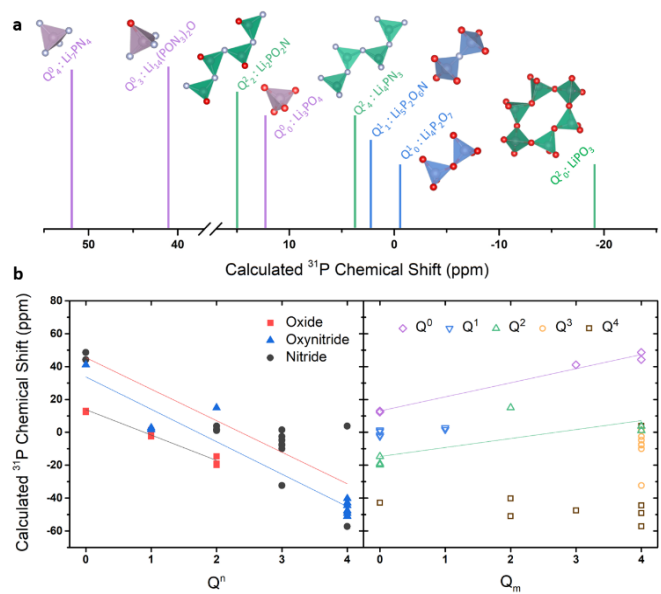

**Figure 2.** (a) Calculated isotropic $^{31}$P chemical shifts of representative $Q^n_m$ structures to illustrate chemical shift variation with Q speciation. (b) Calculated isotropic $^{31}$P chemical shift variation with $Q^n$ (left) and $Q_m$ (right) to reveal the effect of network connectivity and N incorporation, respectively

species. This trend agrees with previous studies investigating the effect of nitridation of phosphate glasses that found N has a deshielding effect on $^{31}$P when replacing O as it has less electronegativity.[26,30,37,38]

A few structures deviate from these $\delta_{iso}$ trends, such as Li$_5$P$_2$O$_6$N, where all $Q^1_1$ units with bridging N is much lower ($\delta_{iso}$=~2 ppm) than the chemical shift for $Q^1_1$ units ($\delta_{iso}$= 14 ppm) found in LiPON. This is due to the strong dependence of the P-N-P bond angle with chemical shift, likely a shielding effect from overlapping terminal P=O bonds that are much weaker in LiPON (SI). The database includes the only two compounds with $N_t$ environments (P$_3$N$_5$ and P$_4$N$_6$O), having $^{31}$P chemical shifts in the range -57 to -44 ppm, well outside the range observed for LiPON. Additionally, no $^{31}$P chemical shift corresponding to Li$_3$P ($\delta_{iso}$=-278 ppm) was observed in the MAS LiPON spectrum, indicating the intermetallic phase does not form as an impurity within or on the surface of the thin film. However, an unknown impurity phase is detected at 115 ppm that is tentatively assigned to three coordinated P defect sites at the surface (SI).

As Fig. 2 shows, many $Q^n_m$ units have isotropic chemical shifts in the range found for LiPON (20-0 ppm), thus comparison of isotropic chemical shifts alone cannot be used for definitive assignments. However, the GIPAW computational method calculates the full chemical shift tensor, which translates to the chemical shift anisotropy (CSA). The CSA reflects the distortion of the electronic structure around the nucleus and contains information regarding the local symmetry of said nucleus, which is fully described by two parameters, anisotropy $\Delta\delta$ and asymmetry $\eta$ (see SI for full convention definition). These CSA parameters provide distinction of different structures based on their local symmetry despite having similar chemical shifts. CSA analysis is especially useful for disordered structures as it provides the means to distinguish different chemical environments despite overlapping resonances.[36,39,40] The calculated $\delta_{iso}$ and CSA parameters are grouped by their corresponding $Q^n_m$ speciation (Fig. 3) to reveal distinct clustering primarily by their $Q^n$ designation. The $Q^0_m$ units are all marked by negligible $\Delta\delta$ and $\eta$ close to 1, reflecting the tetrahedral symmetry of isolated P tetrahedra. The $Q^1_m$ units display moderate and positive $\Delta\delta$ values ranging from 50 to 120 ppm with $\eta$ around 0.4, while the $Q^2_m$ and $Q^3_m$ units have negative and large $\Delta\delta$ values with eta less than 0.4. These CSA differences are key to identifying and correctly assigning the resonances seen in LiPON at higher chemical shifts.

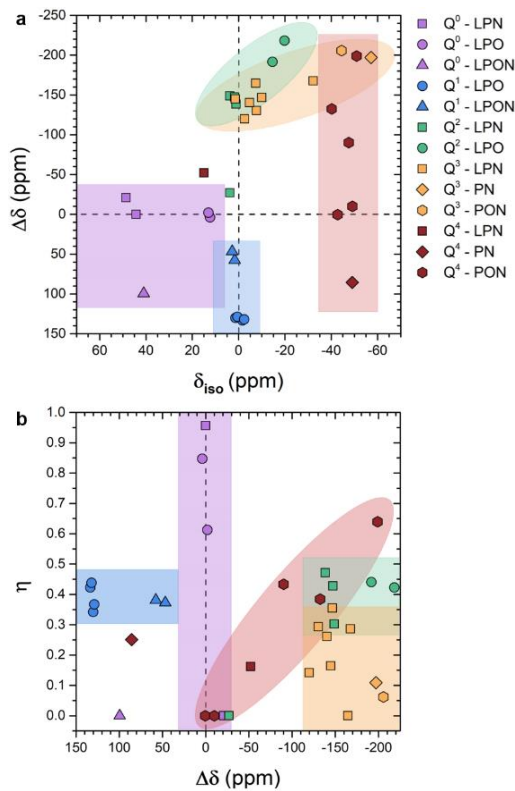

**Figure 3.** Calculated (a) anisotropy-isotropic chemical shift and (b) anisotropy-asymmetry correlation maps of the GIPAW database compounds grouped by $Q^n_m$ speciation showing distinct ranges.

## Computational spectroscopy of LiPON glass

AIMD was used to generate representative model structures from which the chemical shift tensors of the constituent local structural units can be calculated. As there are limited number of oxynitride crystals available, this method removes uncertainty of assignments in the composition gaps in the alkali phosphorus oxynitride variants. To both confirm application of the structural database to the amorphous structure demonstrated in LiPON, AIMD approach is employed to generate an amorphous structure with a stoichiometry of Li$_{2.9}$PO$_{3.5}$N$_{0.31}$, shown in Figure 4a.[15] Previous studies performed AIMD-based melt quenches on a variety of LiPON stoichiometries, clearly linking the Li and N content to the potential for bridging configurations and subsequently improved ionic conductivity, $\sigma_i$, via modified coulombic interactions.[22] They also concluded the low density achieved through AIMD melt/quenching rules out previous interpretations of the opening of the structure to $\sigma_i$ improvement, but correlating $\sigma_i$ to decreased density to improved conductivity.[22]

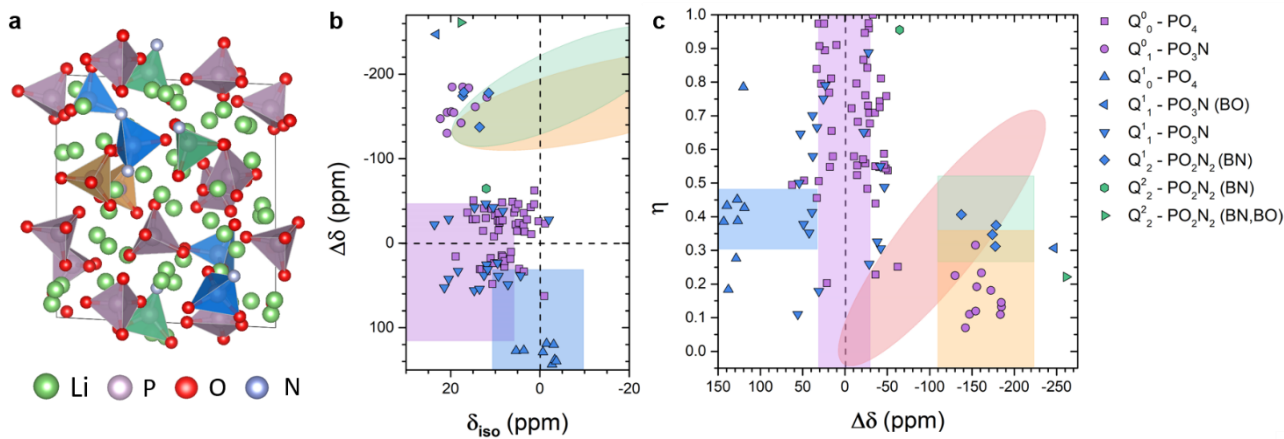

**Figure 4.** (a) Schematic AIMD model of LiPON. Coloration of the P tetrahedra correspond to the $Q^n_m$ speciation. $Q^0_0$ purple, $Q^1_1$ blue, $Q^0_1$ green, $Q^1_0$ orange. (b) Anisotropy-isotropic chemical shift and (c) anisotropy-asymmetry correlation maps of the AIMD model grouped by $Q^n_m$ speciation showing distinct ranges.

However, distinctions between classical metaphosphate glasses and vapor deposited glasses emphasize limitations of the melt-quench method for producing the high-density glasses attained through physical vapor deposition. To emulate the high density of a vapor deposited glass, NVT quenches with densities on the order of crystalline analogues were performed. The AIMD-determined structures are generally consistent with that previously reported, however, upon repeated melts and quenches of the structures of increased density, occasional variations in coordination are observed including the formation of a $Q^2_2$ units in a trimer chain, clearly emphasizing the propensity for N as a bridging unit; the low number of such structures suggests they would be difficult to detect with conventional solid-state NMR techniques.

Using this structure, GIPAW calculations are performed to calculate the relationship between bonding environment and electron shielding (Figure 4b,c). A range of $\delta_{iso}$ are present for the $^{31}$P calculations, likely due to variation in Li coordinations and bond angles. Consistent with observations from the structural database, incorporation of N results in an increased $\delta_{iso}$, whereas the lowest $\delta_{iso}$ is associated with bridging oxygen. While a projection of these datapoints mirror the experimental $\delta_{iso}$ range, calculations of the CSA parameters $\Delta\delta$ and $\eta$ may clearly be used to deconvolute this clustering. For example, $Q^1_0$ units show a distinct $\Delta\delta$ >100 ppm. The average $\delta_{iso}$ and CSA parameters for the corresponding sites are listed in Table 1. The model predicts a structure that is dominated by $Q^0_0$ units followed by $Q^1_1$ and $Q^0_1$ units with minor amounts of $Q^1_0$ (Table 1) and indicates N incorporation prefers bridging over non-bridging sites. The model also predicts a singular $Q^2_2$ unit existing as the center tetrahedra in a trimer chain. Given the limited size of the unit cell and relatively small number of atoms, we do not consider these trimer units to be representative structural units in LiPON.

The crystal database and AIMD model provide important insight regarding the nature of N incorporation in LiPON and related compounds. Despite many claims, primarily spurred by XPS assignments, stating N is coordinated to 3 P tetrahedra ($N_t$) to form a tricluster in LiPON and related phosphorus oxynitride glasses, we find no evidence in the present study to support these assignments, consistent with previous findings using other techniques.[15,21,41] A detailed discussion on this topic with results from GIPAW calculations of $^{15}$N chemical shifts is provided in the SI.

**2D NMR spectroscopy of chemical shift anisotropies**

As the 1D $^{31}$P spectrum (Fig. 1) and the AIMD model (Fig. 4) reveal, there is significant overlap of the constituent $Q^n_m$ $\delta_{iso}$ making deconvolution of the 1D MAS spectrum non-trivial. But as the CSA from the AIMD model shows, there are substantial differences between the CSA of the $Q^n_m$ units, which permits identification of convoluted peaks. The MATPASS/CPMG pulse sequence is used to sequester the anisotropic components into a secondary dimension that can be modelled to extract the CSA parameters at each isotropic chemical shift (SI).[42] Additionally, the projection of the 2D experiment produces a spectrum that is free of CSA, having only isotropic contributions to the chemical shift (Fig. 5 top). The intensity above 20 ppm is not completely refocused due to rapid dephasing of the $Q^0_1$ units from short $T_2$ (see SI). It is evident the magnitude of $\Delta\delta$ does not vary much, as all measured values fall between 65 and 40 ppm. These anisotropy values are rather small and by comparison to some values in Fig. 3 rule out the presence of $Q^2_m$ or $Q^3_m$ units within LiPON. Rather, the small $\Delta\delta$ values indicate the $Q^0_0$ and $Q^1_1$ units dominate the structure as they tend to have smaller $\Delta\delta$. In the case of the $Q^0_0$ units observed in the crystals $\Delta\delta$ is nearly zero, reflecting a symmetric site, whereas in LiPON it is reasonable to consider deviations from this local symmetry arising from a distribution of bond lengths hence the larger $\Delta\delta$. The variations of $\Delta\delta$ and $\eta$ with $\delta_{iso}$ display three regions where the values plateau (Fig. 5b, solid lines), indicating minimal overlap of multiple resonances and the predominance of a singular chemical environment. These plateaus correspond to three of the constituent $Q^n_m$ units: $Q^0_0$ ($\delta_{iso}$ = 9 ppm) with $\Delta\delta$= -42 and $\eta$=0.54, $Q^1_1$ ($\delta_{iso}$ =14) with $\Delta\delta$=63 and $\eta$=0.36, and $Q^1_0$ ($\delta_{iso}$ =3.8) with $\Delta\delta$=-61 and $\eta$=0.06. It should be noted that $\eta$ appears as 1.0 in the case of two superimposed sideband patterns with opposite signs of $\Delta\delta$ and explains rise of $\eta$ at $\delta_{iso}$ = 13.2 and 20.4 ppm; these points are artifacts denoting overlapping regions and produce a gradual rise and fall with $\delta_{iso}$. The gradual changes between the CSA parameters provide guidance on the peak width of their corresponding sites to aid in deconvolution of the MAS spectrum. One component not featured is the $Q^0_1$ peak at 19.4 ppm, as it has greatly lower intensity in comparison to the MAS spectra in Fig. 1. This absence is a result of the rapid dephasing occurring during the MATPASS pulse sequence at lower spinning speeds, making it unable to refocus this component. However, traditional side band analysis at various spinning speeds reveals the CSA of this site to be significantly larger than other sites ($\Delta\delta$ = -150 ppm and $\eta$ = 0.3). The sideband analysis also provides CSA parameters of the other sites that are consistent with the results from MATPASS and the AIMD model. Overall, the MATPASS and sideband CSA analysis indicate there are four peaks: the $Q^0_0$ and $Q^1_1$ sites, having relatively small anisotropies, and the $Q^0_1$ and $Q^1_0$ sites, having much large anisotropies. MATPASS results and a comparison to the MAS sideband analysis is detailed in the supplementary information. The corresponding values agree with the calculated values from the AIMD model.

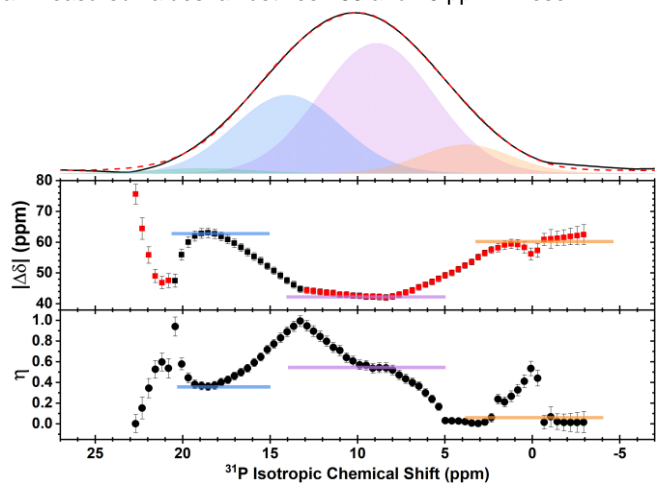

**Figure 5.** $^{31}$P MATPASS (top) Isotropic projection with deconvolution informed by CSA parameter variation, colors are consistent with Fig. 1. (middle) Anisotropy and (bottom) asymmetry variation with isotropic chemical shift. Solid lines denote CSA values reflecting one component.

**Table 1.** NMR parameters used for deconvolution of the 1D MAS $^{31}$P spectrum of LiPON spinning at 25 kHz and the corresponding NMR parameters obtained from the AIMD model of LiPON.

| Site | $\delta_{iso}$ (ppm)[a] | $\delta$ width (ppm)[a] | $\Delta\delta$ (ppm)[b] | $\eta$ | Relative Fraction (%) |
|---|---|---|---|---|---|
| 1D MAS | | | | | |
| $Q^0_0$ | 9.3 | 6.1 | -37 | 0.67 | 49 |
| $Q^1_1$ | 14.6 | 5.8 | 42 | 0.70 | 30 |
| $Q^0_1$ | 19.4 | 6.5 | -150 | 0.30 | 14 |
| $Q^1_0$ | 4.7 | 10 | -98 | 0.30 | 7 |
| AIMD | | | | | |
| $Q^0_0$ | 7.82 | 4.55 | 29.7$^b$ ±12.8 | 0.67 ±0.19 | 53.1 |
| $Q^1_1$ | 12.77 | 6.15 | 38.4$^b$ ±10.1 | 0.50 ±0.21 | 19.8 |
| $Q^0_1$ | 18.15 | 2.96 | -161 ± 17.6 | 0.17±0.07 | 11.5 |
| $Q^1_0$ | -0.75 | 3.19 | 130 ±8.5 | 0.42±0.16 | 8.3 |
| $Q^1_2$ | 14.83 | 2.45 | -166.9±17 | 0.36±0.03 | 4.2 |
| $Q^1_1$ | 23.21 | - | -247 | 0.31 | 1 |
| $Q^2_2$ | 12.04 | - | -64.5 | 0.96 | 1 |
| $Q^2_2$ | 17.72 | - | -261 | 0.22 | 1 |

[a] AIMD shift and width indicate the average and standard deviation of $\delta_{iso}$, respectively. [b] Absolute value for $\Delta\delta$ used as an estimate of the magnitude of the CSA, otherwise underestimated due to sign variation.

In conjunction with the CSA analysis, further insight into the chemical identity and connectivity of the local structure comes from double-quantum (DQ) build-up curves and DQ-SQ correlation spectroscopy (Figure S9). These DQ experiments (detailed in the SI) rely on probing the $^{31}$P-$^{31}$P homonuclear dipolar coupling interaction and can reveal the connectivity of $Q^n$ environments, potentially revealing details on extended chain environments.[43] The results from the buildup curves produce P-P interatomic distances in agreement with those from neutron scattering.[15] DQSQ correlation spectroscopy shows that all P environments are correlated with themselves and all other units indicating the $Q^n_m$ units are randomly distributed through the network. The results also support the identification of the $^{31}$P chemical shift at 19 ppm to the $Q^0_1$ unit. These results solidify there are no extended chain structures or layers within LiPON and indicate the network structure is dominated by isolated P tetrahedra and dimeric units.

The combined experimental and computational results reveal the structure of LiPON is composed of $Q^0_0$ (49%), $Q^1_1$ (30%), $Q^0_1$(14%), and $Q^1_0$ (7%) units, with assignments and isotropic chemical shifts informed by their CSA parameters and comparison to the AIMD model. As the $^{31}$P NMR results were collected quantitatively, the 1D MAS deconvolution can be used to estimate the composition as a check for internal consistency of the assignments. This produces a composition of $Li_{2.93}PO_{3.52}N_{0.30}$, mirroring films of similar ionic conductivity. The AIMD model suggests there may be other minor structural units present making up 1% of the P units, though these are all too low in concentration to observe experimentally and are omitted from the deconvolution. The $Q^1_2$ unit is in slightly higher concentration (~4%) though its $\delta_{iso}$ is expected to be close to the $Q^1_1$ unit and cannot be fully resolved in the spectra; its contribution is included into the $Q^1_1$ peak. Although if this contribution is included separately and makes up to 4% of the P environments a marginally closer estimate of the composition is obtained. Additionally, the deconvolution allows us to indirectly determine the quantity of $N_d$ and $N_a$, as the $N_a$ are exclusively associated with $Q^0_1$ units and all $N_d$ are forming dimers in the $Q^1_1$ units. From our NMR assignments and fit we obtain 67% $N_d$ & 33% $N_a$. These values are entirely consistent with the results obtained by Lacivita et al. who found the $N_d$:$N_a$ ratio depends on the Li:O+N ratio, predicting 60% $N_d$ and 40% $N_a$ for our composition.[15] The internal consistency and agreement of our assignments with other experimental and modelling results indicate our assignments and deconvolution accurately represent the structure of LiPON and its structure is unequivocally solved.

### A Glassy Perspective of LiPON

Understanding the structure of LiPON is essential for determining its electrochemical properties. The general trend for lithium phosphorus oxynitride glasses is with higher Li content the conductivity concomitantly increases, as expected from $\sigma_i = ne\mu$. However, at high Li content, N incorporation is shown to enhance the mobility through formation of dimers with bridging N which attract Li less strongly than PO bonds.[15,22] There is a limit however, as increasing Li relative to N breaks the $N_d$ to form non-bridging N, that have a stronger interaction with Li and consequently lowers conductivity. The structural assignments developed here indicate [31]P NMR can be used as an indirect measure of the $N_a$ to $N_d$ ratio in LiPON related materials. The introduction of N into the network however does not necessarily provide any indication as to why LiPON is so stable against Li metal.

Considering LiPON is a glassy material, we advocate a possible theory for LiPON's stability that relies on its glassy nature.[44] In accordance to the ultra-stable glasses investigated by Ediger et al., physical vapor deposited (PVD) glasses show remarkably low fictive temperatures indicating they are close to the bottom of their potential energy landscapes, resulting in kinetic and chemical stabilities that cannot be achieved by conventional heat treatments on reasonable timescales.[45,46] This enhanced kinetic stability has been observed in PVD organic[47], metallic[48], and chalcogenide[49] glasses and considering LiPON is grown by a form of PVD, it is reasonable to assume it too can display a low fictive temperature after deposition. The implication of LiPON as a low fictive temperature glass is that it is nearing the bottom of its potential energy landscape thus the energy difference between the metastable glassy state and the corresponding crystalline state is minute.[50] This minimizes the thermodynamic driving force for crystallization and the enthalpy barriers for initiating structural rearrangement are too high to overcome on an experimental timescale and are consequently suppressed. This enhanced kinetic stability observed in ultra-stable glasses could be a possible explanation for the superior electrochemical stability LiPON presents with Li metal. Even with the interfacial driving force to decompose, the kinetic stability of LiPON in its 'stable' form may reduce the decomposition rate. This also implies that LiPON's stability may not be only related to the chemistry of LiPON but also a consequence of the unique synthesis route in which it is made. How ultra-stable glass kinetic stability relates to electrochemical stability remains to be seen and requires further studies explicitly investigating the connection. Previous work has explored the increased kinetic stabilization of LiPON via annealing at a variety of temperatures below the measured $T_g$ of bulk counterparts.

Among their results, annealing temperature was shown to have little effect on composition, attributing all conductivity changes to structural and configurational modifications, albeit in 50 nm thick films.[51] While deposition is not controlled in these experiments, literature has reported an increase in temperature up to 110°C due to plasma heating of the film during deposition.[52] Such surface heating enhances mobility of surface ions, resulting in increased glass density.[47] However, this is below the critical annealing temperature before a severe drop of ionic conductivity is observed (~150°C).[44] The potential for plasma heating is one further variable among the LiPON deposition field, and likely accounts for variable performance and stability.

Last, sources of the high degree of cyclability of Li/LiPON cells likely extend beyond electrochemical stability of LiPON itself. Mechanically, the lack of connectivity coupled with the high cation concentration appears to manifest a high degree of film compliance. As the structure is dominated by $Q^0_0$ and $Q^1_1$ units with N acting to bridge about 30% of the phosphate tetrahedra as dimeric units, the overall LiPON network structure clearly does not contain any extended chain structures as indicated by DQSQ results and absence of CSA values expected for $Q^2$ environments. Interestingly, the free-standing films of LiPON produced for this study exhibit a high degree of flexibility considering the film thickness (Figure 6b). This degree of compliance is surprising, and questions current requisites for a solid state electrolyte to resist dendrite penetration, generally purported to require a critical modulus.[53] Such flexibility is commonly observed in chemically-tempered alkali-aluminosilicate glasses, where fracture is prevented by a lack of surface defects. The flexibility in LiPON glass suggests that the presence of undercoordinated P groups (see SI) do not behave as defects, or as a detriment to the film's mechanical properties. Such flexibility is likely enhanced by using PVD processes, which produces smooth, uniform films. This mechanical compliance will be explored in future work.

## Conclusion

Using 1D and 2D solid-state NMR methodologies, the local structure of amorphous LiPON is definitively resolved, showing the prevalence of $Q^0_0$ tetrahedra and identifying N incorporation to form dimeric units via bridging N and separately non-bridging N on orthophosphate tetrahedra. GIPAW methodologies permit calculation of a range of phosphate-based compounds, clearly identifying trends in chemical shift tensors as a function of composition and local structure. Fitting of chemical shift anisotropy parameters of LiPON are determined by combining AIMD to amorphize of LiPON and GIPAW methodologies for calculation of the electronic shielding associated with chemical shifts observed in NMR. The high stability of LiPON is described structurally as a combination of the low connectivity of the structure as well as the hyperannealing that occurs with physical vapor deposition. Free-standing films of LiPON are produced, exhibiting a high degree of flexibility, and hence compliance, which further supports the lack of long-range order and questions the role of mechanical properties in the cyclability of LiPON.

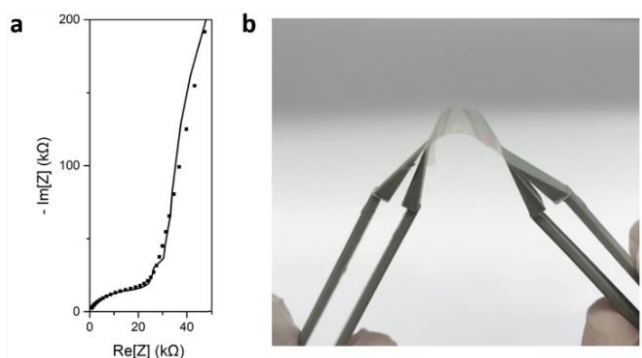

**Figure 6.** A 3.75 µm thick free-standing film of LiPON is produced, showing (a) ionic conductivity similar to its confined film counterparts ($\sigma_i$ = 2.6 µS/cm) and (b) a remarkable degree of film compliance.

## Acknowledgements

The authors gratefully acknowledge funding support from the U.S. Department of Energy, Office of Basic Energy Sciences, under Award Number DE-SC0002357 (program manager Dr. Jane Zhu). XPS were performed at the UC Irvine Materials Research Institute (IMRI). XPS work was performed at the UC Irvine Materials Research Institute (IMRI) using instrumentation funded in part by the National Science Foundation Major Research Instrumentation Program under grant no. CHE-1338173. This work also used the Extreme Science and Engineering Discovery Environment (XSEDE), which is supported by National Science Foundation grant number ACI-1548562. This work was supported by Laboratory Directed Research and Development (LDRD) Award 20-FS-012 and performed under the auspices of the U.S. Department of Energy by Lawrence Livermore National Laboratory under Contract DE-AC52-07NA27344. LLNL-JRNL-810272-DRAFT

**Keywords:** NMR spectroscopy • Ab initio calculations • Solid Electrolyte • GIPAW • LiPON

## Table of Contents

# SUPPORTING INFORMATION

## Experimental Procedures

### Thin film deposition

4µm thick thin films of a-Li$_3$PO$_4$ and LiPON were deposited via RF-sputtering. Films were deposited from a 5 cm c-Li$_3$PO$_4$ target at a power of 50W with minimal reflected power (~1W) in Ar (15 mTorr) and N (15m Torr) gas. 98% enriched $^{15}$N gas was incorporated into the film at a ratio of 1:4 relative to standard ultra-high purity N$_2$. To maximize surface area to minimize signal-to-noise, films were deposited on both sides of cylindrical quartz rods or in tested in free-standing form. Synthesis of the free standing LiPON will be described further in an upcoming manuscript.

### NMR Spectroscopy

The magic angle spinning (MAS) NMR measurements performed on the Li$_3$PO$_4$ target material, amorphous Li$_3$PO$_4$ film, and LiPON films deposited on fused quartz rods were collected using a 4 mm X/H channel Revolution probe on a 400 MHz (9.4 T) Bruker Biospin Avance Neo, operating at 161.97, 155.50, and 58.88 MHz for $^{31}$P, $^7$Li, and $^6$Li, respectively. The samples were packed within a 4mm pencil-type ZrO$_2$ rotor and spun at 10 kHz. The $^{31}$P and $^6$Li spectra were collected as a rotor synchronized Hahn echo experiment with a 90° and 180° pulse of 2.4 and 4.8 µs, respectively (B$_1$ field strength ~104 kHz) for $^{31}$P and 4.25 and 8.5 µs pulse lengths for $^6$Li. The Hahn echo experiments were processed from the top of the echo to remove the effects of ring down from the FID. A single pulse experiment with a pulse length of 2.875 µs (B$_1$ field strength ~87 kHz) was used to acquire the $^7$Li spectra. The T$_1$ for $^{31}$P and $^7$Li were measured with inversion recovery and found to be 6.5 s and 332 ms, respectively. The recycle delays used for the 1D experiments were 25 s, 2 s, and 30 s for $^{31}$P, $^7$Li, and $^6$Li, respectively.

The $^{31}$P magic angle turning phase adjusted sideband separation (MATPASS)/CPMG experiment utilizes five π pulses of length 4.8 µs with rotor synchronized inter-pulse delays following those outlined by Hung et al.[1] The spin speed for the experiment was 4 kHz and for 2D acquisition 16 hypercomplex t$_1$ points were collected with 1008 transients per t$_1$ point and 22 echoes per transient with a recycle delay of 19.5 s. The hypercomplex data was acquired with the method of STATES [2] employed for the phases of the receiver and CPMG pulses. The 2D MATPASS/CPMG dataset was processed using a custom Python script that follows typical processing steps outlined by Hung et al.[1] along with fitting the T$_2$ decays from the CPMG echo train and adjusting their spectral components through CPMG reconstruction.[3,4] The implementation of the CPMG reconstruction processing with the MATPASS/CPMG pulse sequence will be discussed in a forthcoming publication.

The MAS NMR measurements performed on the free standing LiPON film were collected using a 2.5 mm triple resonance Bruker probe on a 600 MHz (14 T) Bruker Biospin Avance III, operating at 242.94 and 233.23 MHz for $^{31}$P and $^7$Li, respectively. Sheets of the free-standing film were stacked and lightly tamped within a 2.5 mm ZrO$_2$ rotor and spun at 25 kHz. The $^{31}$P spectrum was collected as a rotor synchronized Hahn echo experiment with a 90° and 180° pulse of 4.45 and 8.9 µs, respectively (B$_1$ field strength ~112 kHz) for $^{31}$P. The Hahn echo experiment was processed from the top of the echo. The slow spinning speed $^{31}$P MAS spectra (5, 10, 15 kHz) were collected with a single pulse experiment with the same 90° pulse length as the Hahn echo experiment. The 1D $^{31}$P MAS measurements had a recycle delay of 20 s. Additionally, a $^{31}$P CPMG echo train was recorded to determine T$_2$ which was found to be 3.355 ms. A single pulse experiment with a pulse length of 2.875 µs (B$_1$ field strength ~87 kHz) was used to acquire the $^7$Li spectra with a recycle delay of 2 s.

The $^{31}$P-$^{31}$P homonuclear double quantum experiments were collected with a 'back-to-back' BaBa-xy16 pulse sequence[5] while spinning at 25 kHz and using a 90° pulse length of 4.45 µs. The DQ build-up curves were collected by recording a double quantum filtered (DQF) spectra and a reference spectrum, having the same experimental conditions except for a phase shift of the reconversion period, at progressively larger excitation/reconversion times incremented by 4×N$T_R$, full rotor period. Each DQF and reference spectrum collected 64 transients and had a 20 s recycle delay. The double quantum single quantum (DQSQ) correlation experiment used an excitation and reconversion time of 0.72 ms (18 rotor periods). The 2D DQSQ experiment was acquired with 64 t$_1$ points, with 256 transients per t$_1$, using a rotor-synchronized t$_1$ increment of 40 µs and 20 s recycle delay. Processing of the DQSQ correlation contour plot was performed using the software package ssNake.[6]

All $^{31}$P and $^{7/6}$Li spectra were externally referenced to hydroxyapatite ($\delta_{iso}$=2.65 ppm relative to 85% H$_3$PO$_4$ in H$_2$O) and 1M LiCl (aq) solution ($\delta_{iso}$=0 ppm), respectively. Deconvolution of the $^{31}$P MAS line shapes were carried out using the software package dmfit.[7] Analysis of the F1 dimension of the MATPASS data for determination of chemical shift anisotropy parameters was carried out with the software HBA.[8] The $^{31}$P sideband analysis of the variable MAS of the FS LiPON film was carried out using the software package dmfit.[7] The principle components of the chemical shift tensor ($\delta_{xx}$, $\delta_{yy}$, $\delta_{zz}$) are expressed in accordance with the Haeberlen convention by the isotropic chemical shift $\delta_{iso}$, magnitude of anisotropy Δδ, and asymmetry parameter η.[9] These parameters are defined as:

$$\delta_{iso} = \frac{1}{3}(\delta_{xx} + \delta_{yy} + \delta_{zz})\,;\; \Delta\delta = \delta_{zz} - (\delta_{xx} + \delta_{yy})/2\,;\; \eta = \frac{\delta_{yy}-\delta_{xx}}{\delta_{zz}-\delta_{iso}}$$

where the principle components are ordered by $|\delta_{zz} - \delta_{iso}| \geq |\delta_{xx} - \delta_{iso}| \geq |\delta_{yy} - \delta_{iso}|$.

### Density functional theory

DFT calculations were performed using the Vienna *ab initio* Simulation Package (VASP).[10] Structures were extracted from either the Materials Project[11] or the International Crystal Structure Database[12], when available. GIPAW[13] calculations were performed using the PBE functional, and a plane-wave basis set cutoff energy of 800 eV. Automatic generated gamma-centered k-points grids in excess of 1,500 points/A$^3$ were determined to converge CSA results. AIMD was performed using a Langevin thermostat in NPT for the case of a standard glass quench[14] and NVT mode for simulation of the hyperannealed glass, heating structures with reduced unit cell volume (10% lower than NPT quenched structures) to 3000 K and cooling at a linear rate of 3 K/ps. Stoichiometries are selected based on experimental methodologies, and quenched structures are confirmed to be consistent with those of the similar stoichiometry in literature.[15]

### X-ray photoelectron spectroscopy (XPS)

XPS was conducted using a Kratos AXIS Supra with the Al anode source operated at 15 kV with a 500 mm Rowland circle monochromator. The chamber pressure was <10$^{-8}$ Torr during all measurements. High resolution spectra were calibrated using the hydrocarbon C1s peak at 284.8 eV. Fitting was conducted using CasaXPS software using a Shirley-type background. Samples were transferred to the XPS chamber from a glove box via vacuum transfer. All peaks were fit with a GL(30) Gaussian Lorentzian line shape and the peak positions are outlined in Table S1.

### Electrochemical impedance spectroscopy (EIS)

EIS was performed using a Biologic SP-200 on deposited films with Cu electrodes, modulating a potential of +/-10mV at frequencies ranging from 500 mHz to 3MHz, sampling 6 frequencies per decade on a logarithmic scale. 5 measurements were performed at each frequency and averaged. Fitting was performed using Biologic ECLab software.

## Results and Discussion

### Baseline properties of thin film amorphous LiPON

Free-standing LiPON thin films were tested for ionic conductivity and the nature of bonding using electrochemical impedance spectroscopy (EIS) and x-ray photoelectron spectroscopy (XPS), respectively. EIS produced an ionic conductivity of $2.6\times10^{-6}$ S/cm, on the order of those produced in literature via RF sputtering (EIS in Fig. 6a in the main text, circuit element in Fig. S2, fit in Table S2). XPS results are consistent with literature, showing minimal surface contamination, viewed through the C intensities, and bonding consistent with films previously reported (Fig. S1). However, per recent literature discussion,[16] the peak observed around 399 eV is attributed to the apical nitrogen sites, rather than the commonly interpreted $N_t$. Fit parameters of the XPS data is shown in Table S1.

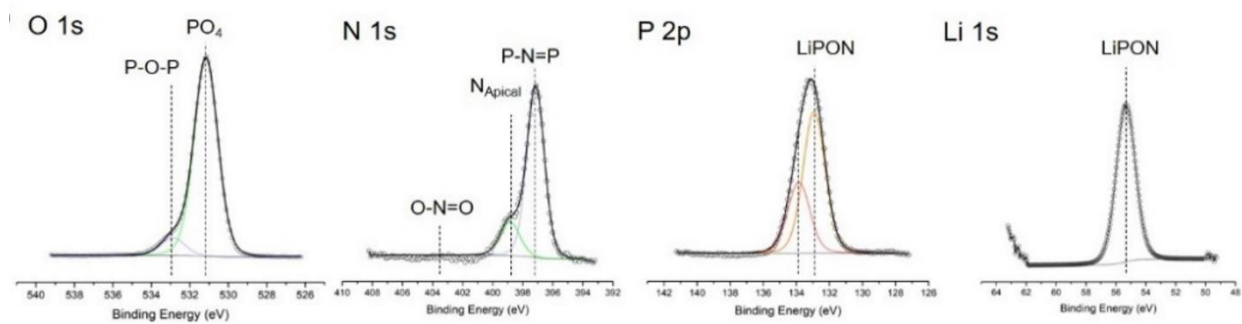

**Figure S1.** Baseline properties of LiPON determined by (a) electrochemical impedance spectroscopy and (b) x-ray photoelectron spectroscopy.

**Table S1.** Fit for LiPON XPS spectra

|  | Li 1s | O 1s | P 2p | N 1s | C 1s |
|---|---|---|---|---|---|
| LiPON | 55.1 |  | 133.0 |  |  |
|  | 55.0 | 55.6 | 132.8[17] |  |  |
|  | 55.4[18] | 55.8 | 133.6 |  |  |
| P-O-P |  | 532.6 |  |  |  |
|  |  | 532.7   532.8 |  |  |  |
|  |  | 533.0[19]   532.3[20] |  |  |  |
| $PO_4$ |  | 531.0 |  |  |  |
|  |  | 530.6[21]   531.3[20] |  |  |  |
|  |  | 513.4[19] |  |  |  |
| O-N=O |  |  |  | 403.1 |  |
|  |  |  |  | 404.0[22] |  |
| P-$N_a$ |  |  |  | 398.7 |  |
|  |  |  |  | 398.6[21]   399.0[19] |  |
|  |  |  |  | 398.9[22]   399.4[20] |  |
| P-N=P |  |  |  | 397.2 |  |
|  |  |  |  | 396.6[21]   397.6[19] |  |
|  |  |  |  | 397.4[22]   397.8[20] |  |
| $Li_2CO_3$ | 55.3 | 531.6 |  |  | 298.6 |
|  | 55.3[23] | 531.9[23] |  |  | 290.1[23] |
|  | 55.2[24] | 531.5[24] |  |  | 289.8[24] |
| LiOH | 54.6 | 531.1 |  |  |  |
|  | 54.7[23] | 531.1[23] |  |  |  |
|  | 54.9[24] | 531.3[24] |  |  |  |
| $NH_3$ |  |  |  | 398.6 |  |
|  |  |  |  | 398.5-400.9[25,26] |  |

# SUPPORTING INFORMATION

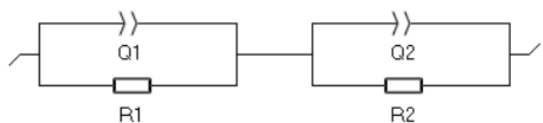

**Figure S2.** Equivalent circuit of freestanding LiPON.

**Table S2.** EIS fitting parameters

| Q1 | a1 | R1 | Q2 | a2 | R2 |
| --- | --- | --- | --- | --- | --- |
| 1.79E-10 | 0.8787 | 26779 | 3.54E-10 | 0.9765 | 3.11E+06 |

**SUPPORTING INFORMATION**

**NMR spectra of Li$_3$PO$_4$ target and amorphous thin film**

For a reference, $^{31}$P NMR was collected on the target material crystalline Li$_3$PO$_4$ as shown in Figure S3a. The spectrum shows a narrow peak at 9.6 ppm that is assigned to a Q$^0$ environment, PO$_4^{3-}$ tetrahedra, as expected from the crystal structure. However, a small signal is observed around -5 ppm, likely a result of decomposition to form triclinic and pseudo-monoclinic pyrophosphate phases, Li$_4$P$_2$O$_7$, upon exposure to moisture.[27] For a more accurate comparison to LiPON films and to understand how the incorporation of N influences the $^{31}$P spectrum, amorphous Li$_3$PO$_4$ films were grown under similar conditions in Ar rather than N$_2$ gas and the $^{31}$P spectrum was acquired (Fig. S3b). The isotropic chemical shift for Q$^0$ remains nearly the same as the Li$_3$PO$_4$ crystal however the a-Li$_3$PO$_4$ film is broadened in comparison due to structural disorder. A shoulder is observed at -0.4 ppm and is attributed to Q$^1$ species, dimeric phosphate tetrahedra P$_2$O$_7^{4-}$. The rise of Q$^1$ units suggests some Li is lost during the deposition process, assuming a binary model for Q speciation holds for the orthophosphate region[28], then it is estimated about 3 mol% Li$_2$O is lost.

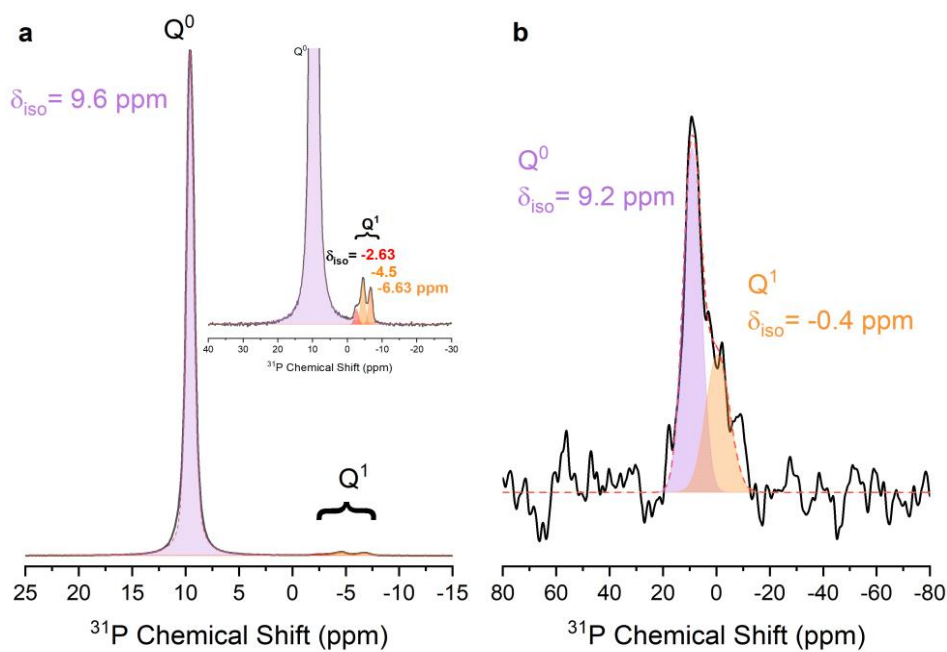

**Figure S3.** 31P NMR spectra of (a) β-Li$_3$PO$_4$ target, inset of magnified region with Q$^1$ decomposition impurities and (b) amorphous Li$_3$PO$_4$ thin film.

**$^6$Li NMR measurements**

While $^7$Li is the more abundant nuclei, it can be a poor nucleus for spectroscopic studies due to dipolar and quadrupolar broadening and exhibits a limited chemical shift range. On the other hand, $^6$Li is largely free from these broadening mechanisms making it a more sensitive probe for changes in Li chemical environments at the expense of lower abundance. A comparison between crystalline $Li_3PO_4$ and LiPON shows there is a similar difference of chemical shift as seen in the $^7$Li spectra as N deshields the Li ions (Fig. S4). Notably, the incorporation of nitrogen into LiPON results in a broader $^6$Li peak, associated with a greater variety of chemical environments for Li ions likely a result from structural disorder. The higher chemical shift with nitridation has been observed before in lithium metaphosphate glasses as well, interpreted as a lowering of the average coordination number of Li.[29]

The measured chemical shift for $^6$Li in LiPON is 1.17 ppm, 0.95 ppm higher than that of crystalline $Li_3PO_4$; a similar difference observed for $^7$Li in LiPON and crystalline $Li_3PO_4$ (0.80 ppm). GIPAW calculations predict average chemical shieldings of -57.45 and -57.22 ppm for LiPON and $Li_3PO_4$, respectively. Like experimental spectra, simulated LiPON exhibits a wide range of chemical shielding values from -55.9 to -58 ppm. The difference of the calculated chemical shieldings between LiPON and $Li_3PO_4$ extends to 1.32 ppm, covering the difference observed experimentally. These values are not corrected relative to the experimental data available, as the mobile Li causes chemical exchange between various sites that experimentally obscure the resonances from being observed. Converting shielding to shift is then made more complicated without thorough experimental results carried out on $^7$Li spectra within the rigid lattice regime and account for the quadrupolar coupling constants. This is level of rigor is not typically carried out for $^7$Li MAS experiments so we have left the calculated shielding values unaltered.

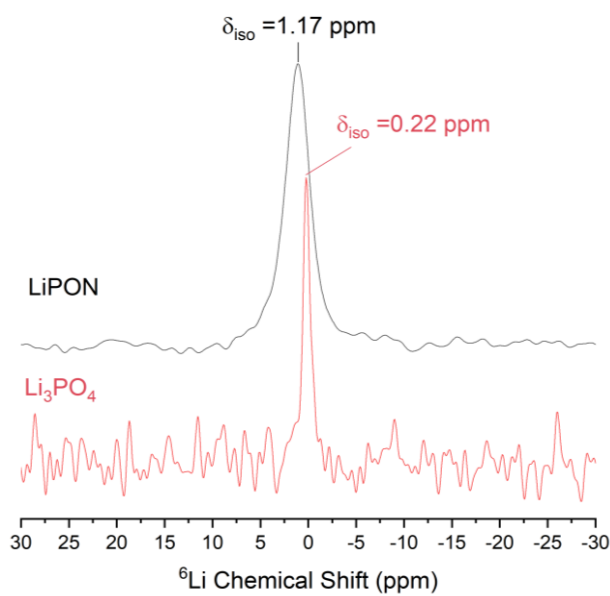

**Figure S4.** $^6$Li NMR comparison between $Li_3PO_4$ and LiPON.

**A Qualitative Look at Thin Film LiPON NMR Signal Enhancement**

Sample limitation is one of the major challenges of collecting NMR signal of LiPON synthesized as a thin film. Fortunately, $^{31}$P and $^{7}$Li are nearly 100% abundant nuclei and make up a considerable amount of the composition of LiPON allowing signal to be detected despite the small quantity of sample. First, a $^{31}$P spectrum was collected on 2 rectangular slabs of LiPON deposited on $Al_2O_3$ substrates placed within the 4 mm $ZrO_2$ rotor and cushioned with PTFE tape. The 5 μm thick LiPON layer coated an area approximately 2 mm wide and 9 mm tall on the slabs. The $^{31}$P spectra was collected for ~20,000 scans (3 days) (Fig. S5, bottom) and has moderate signal to noise. However, the long acquisition time prevents any 2D NMR experiments from being performed within a reasonable amount of time. In order to increase the acquisition rate, we deposited LiPON over 0.5 mm diameter $SiO_2$ rods in order to increase the surface area by approximately a factor of 20. The rods were bundled together tightly and placed within the 4 mm $ZrO_2$ rotor, and signal was clearly collected after 8 scans. A relatively high-resolution spectrum was obtainable after 4,000 scans (16 hours). While increasing acquisition rate by increasing the surface area of the substrate and thus the deposited LiPON proved successful, the acquisition rate is still too slow to perform some 2D NMR measurements like DQ which can have poor recoupling efficiency. Rapid acquisition (8 scans) and excellent signal to noise ratio was only achieved after packing sheets of free standing (FS) LiPON film into a 2.5 mm $ZrO_2$ rotor. The increased sensitivity was a result of the increased fill factor of the sample relative to previous measurements and the decreased distance of the sample to the probe solenoid. Within the resolution obtained of the respective spectra, they display the same peak width and have nearly identical line shapes. This indicates that the FS LiPON film is not altered in any significant way during processing and is the same as LiPON deposited on a substrate.

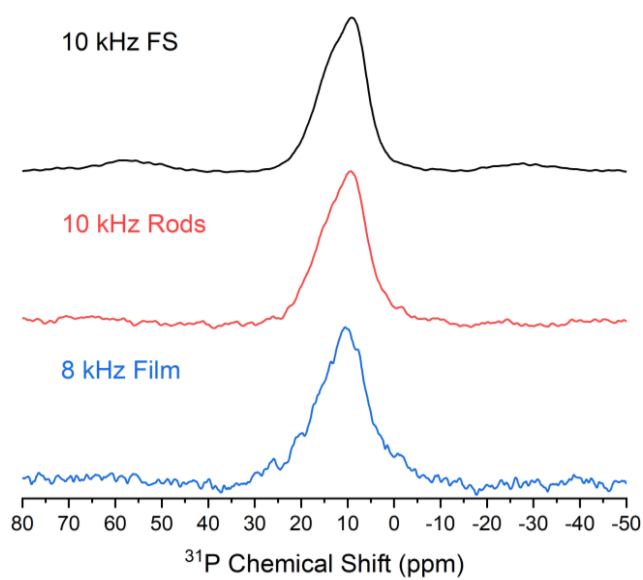

**Figure S5.** Comparison of $^{31}$P spectra of LiPON collected as a single thin film on $Al_2O_3$ substrate (blue), thin film deposited on a bundle of $SiO_2$ rods (red), and a free-standing thin film.

### $^{31}$P CSA MATPASS and Sideband Analysis Results

The compounds in the crystalline database (Fig. 3) reveal there is clustering of resonances between 20 and 0 ppm from oxide, oxynitride, and nitride compounds all with very different local structure, $Q^n_m$, where n refers to the Q speciation and m the number of N coordinated to the measured P. The majority of LiPON intensity resides in this region, meaning comparison of isotropic chemical shifts between known crystals to this amorphous system is limited, especially if the amorphous network contains $Q^n_m$ species not found in the crystalline structures. Hence, we turn to analyzing the chemical shift anisotropy (CSA) to gain further insight for distinguishing and assigning the constituent local structural units in LiPON. The chemical shift is a second rank tensor reflecting the distortion from a spherical distribution of the electronic structure around the nucleus whereas the isotropic chemical shift is the average of the principle components. Thus, while the isotropic chemical shift gives information of the chemical environments, the CSA is more sensitive to changes in the local structural symmetry. The CSA is typically measured by analysis of the side band patterns in slower spinning MAS measurements; however, it can be complicated in the case of overlapping isotropic peaks like those found in glasses. The intensity in the side bands reduces the isotropic intensity creating a situation where either a high resolution isotropic line shape is collected at fast spinning speeds at the cost of losing all CSA information or the spectrum is collected at slower spinning speeds to retain the CSA information with much lower signal to noise. 2D NMR methods like magic-angle turning and phase adjusted sideband separation (MATPASS) can resolve these issues by sequestering the CSA into a second dimension while retaining the isotropic resolution and allows for more discrete analysis of how the CSA varies with isotropic chemical shift.[1,30–32] This method is especially useful for disordered structures as it provides the means to detect different chemical environments within overlapping resonances and gives guidance on peak overlap for deconvolution.[1,30,33]

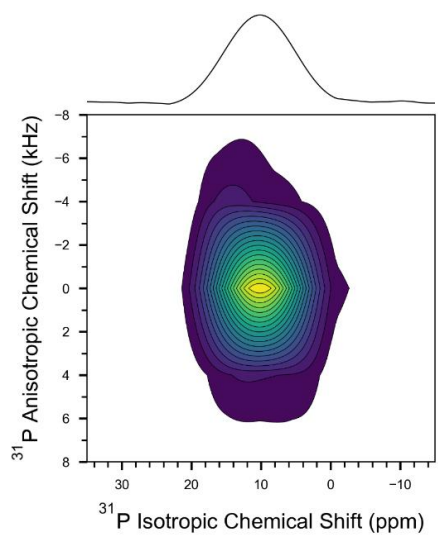

**Figure S6.** 2D $^{31}$P MATPASS contour plot of LiPON showing the CSA along the F1 dimension with the isotropic projection above.

The MATPASS experiment was performed on LiPON deposited on $SiO_2$ rods. The projection of the $^{31}$P MATPASS contour (Fig. S6, top) is a line shape free of any anisotropic broadening and analogous to an infinite spinning speed spectrum. Taking a slice along the anisotropic dimension at an isotropic chemical shift provides a sideband pattern that can be modelled to extract out the CSA parameters Δδ and η, following the Haeberlen convention for describing the chemical shift tensor. It should be noted that because the CSA for the $^{31}$P sites are very small, only 1$^{st}$ order sidebands contain significant intensity adequate for modelling. Attempts at performing the experiment at slower spinning speeds (<3.3 kHz) in order to increase the number of sidebands in the anisotropic dimension were unsuccessful because the inter-pulse delays of the rotor synchronized pulse sequence below 3.3 kHz (2.72 ms) approach or exceed the $^{31}$P $T_2$ of LiPON (3.36 ms). The shown contour plot is then a compromise of spinning fast enough to refocus the signal intensity while spinning slow enough to retain some of anisotropic intensity. Despite these efforts, one site ($δ_{iso}$ =19.4 ppm) is not completely refocused and is largely absent in the MATPASS experiment as discussed below.

By fitting each anisotropic slice and fitting the sideband pattern, the CSA parameters are extracted at each $δ_{iso}$ and plotted in Fig. 5. This analysis reveals there are variations in the CSA parameters despite the mostly featureless isotropic projection. As mentioned in the main text, three plateaus can be identified as having minimal overlap of neighboring resonances with the regions in between these plateaus showing a nearly continuous change in the CSA parameters. These latter transitionary regions represent overlapping resonances as the CSA parameter is a superposition of the two anisotropic line shapes and changes gradually towards the plateau value as one site becomes more dominant. These transitionary regions provide guidance on the peak widths used in the deconvolution of the isotropic projection (Fig. 5) and for the MAS spectra (Fig. 1) by informing where the influence of one site is nearly eliminated. This transition can be seen clearly in Fig. 5 between 19 to 8 ppm, where the Δδ and η start at plateau values of 62 ppm and 0.4 for the $Q^1_1$ site and with decreasing $δ_{iso}$ both values slowly change as the CSA values for the $Q^0_0$ site become dominant until there is minimal signal from the $Q^1_1$ site at 8 ppm. As discussed in the main text, the three plateaus observed for the CSA parameters correspond to the following sites $Q^0_0$ ($δ_{iso}$ = 9 ppm) with Δδ= -42 and η=0.54, $Q^1_1$ ($δ_{iso}$ =14) with Δδ=63 and η=0.36 , and $Q^1_0$ ($δ_{iso}$ =3.8) with Δδ=-61 and η=0.06. These sites CSA values and $δ_{iso}$ are consistent with the corresponding sites produced by the AIMD simulated structure, and support the assignments are correct. Further justification comes from comparing the sites CSA and $\delta_{iso}$ values to those found in crystalline structures (Table S5).

**Table S3.** CSA fitting parameters obtained for the slow MAS experiments (Figure S8) and the plateau CSA values obtained from MATPASS along with the deconvolution parameters for the isotropic projection (Figure 5, top).

| Site Assignment | $\delta_{iso}$ (ppm) | δ width (ppm) | Δδ (ppm) | η | Relative Fraction (%) |
|---|---|---|---|---|---|
| 5 kHz | | | | | |
| $Q^0_0$ | 8.83 | 6.9 | 37.5 | 0.68 | 49.5 |
| $Q^1_1$ | 14.66 | 6.5 | 42 | 0.7 | 29.1 |
| $Q^0_1$ | 18.27 | 7.5 | -150 | 0.3 | 11.9 |
| $Q^1_0$ | 0.69 | 7 | -97.5 | 0.3 | 9.5 |
| 10 kHz | | | | | |
| $Q^0_0$ | 8.63 | 6.8 | -46.5 | 0.45 | 49.4 |
| $Q^1_1$ | 14.41 | 7 | 45 | 0.96 | 30.1 |
| $Q^0_1$ | 19 | 9 | -145.5 | 0.3 | 15.4 |
| $Q^1_0$ | 0.69 | 7 | -97.5 | 0.3 | 5.1 |
| 15 kHz | | | | | |
| $Q^0_0$ | 8.74 | 6.8 | -46.5 | 0.45 | 48 |
| $Q^1_1$ | 14.41 | 7 | 55.5 | 0.45 | 28 |
| $Q^0_1$ | 18.39 | 10 | -145.5 | 0.3 | 19 |
| $Q^1_0$ | 1.5 | 9 | -120 | 0.3 | 5 |
| 25 kHz | | | | | |
| $Q^0_0$ | 9.28 | 6.07 | 46.5 | 0.67 | 48.6 |
| $Q^1_1$ | 14.63 | 5.8 | 40.5 | 0.5 | 30.4 |
| $Q^0_1$ | 19.44 | 6.5 | -150 | 0.15 | 14.5 |
| $Q^1_0$ | 4.7 | 10 | -97.5 | 0.3 | 6.5 |
| MATPASS | | | | | |
| $Q^0_0$ | 8.9 | 7.7 | -42 | 0.54 | 56.6 |
| $Q^1_1$ | 14 | 7.18 | 63 | 0.36 | 31.8 |
| $Q^0_1$ | 18.8 | 6.5 | - | - | 1.8 |
| $Q^1_0$ | 3.82 | 6.08 | -61 | 0.06 | 9.8 |

The $Q^0_0$ site can be directly compared to the $Li_3PO_4$ crystal (both β and γ phases). First the $\delta_{iso}$ of the $Q^0_0$ site in LiPON is found to be 9.3 ppm which is very close to the experimentally observed $Q^0$ value of 9.6 ppm for crystalline β-$Li_3PO_4$ (Fig. S3). The calculated CSA parameters of the two crystals however have Δδ values very close to 0 ppm and η values greater than 0.6 (Table S3); these values reflect the nearly completely symmetric site of $Q^0$ in the $Li_3PO_4$ crystals hence the Δδ value being almost zero. In LiPON, the $Q^0_0$ sites have larger Δδ (40 ppm) due to the distribution of bond lengths inherent from the structural disorder, while on the other hand the η value (0.6) agrees well with the value found for β-$Li_3PO_4$. For comparison of the $Q^1_1$ site ($\delta_{iso}$ = 14 ppm, Δδ = 63 ppm, η=0.36 ), the CSA parameters, including the sign of Δδ, are very close to those found in $Li_5P_2O_6N$ (Δδ =58 or 47 ppm, η = ~0.38), whose structure consists of $PO_3N$ dimers with corner sharing N ($Q^1_1$), indicating the local structure of $Q^1_1$ sites in LiPON are similar to that in $Li_5P_2O_6N$. However, the slightly higher Δδ in the glass reflects some structural disorder of the bond lengths in LiPON. The $\delta_{iso}$ values (~14 ppm) found within the experimental and AIMD LiPON do not agree with the value calculated for $Li_5P_2O_6N$ ($\delta_{iso}$ = ~2 ppm). While the experimental $^{31}P$ shift for $Li_5P_2O_6N$ has yet to be measured (or as far as the authors are aware, has yet to be synthesized and have its structure refined), this discrepancy is surprisingly large. Consideration of the P-N-P bond angles in the AIMD LiPON reveals a dramatic change in the chemical shift as the angle decreases (Fig. S7). The average bond angle in LiPON is found to be 119° whereas in the $Li_5P_2O_6N$ the angle is 125°. An explanation for this dependence could be from the orientation of the double bonds (either P=N or P=O) of the tetrahedra. The electronic density around the double bond is quite large and anisotropic relative to the other P-O bonds and its orientation relative to the P nuclei will have a large effect on shielding or deshielding the P nuclei, hence the broad range observed in Fig. S7. It is possible that as the P-N-P angle decreases the double bond changes orientation relative to the opposite P tetrahedra to minimize electron overlap, consequently increasing chemical shift. The MATPASS derived CSA values for the final site shows Δδ = -61 ppm and η = 0.06. This site was assigned to $Q^1_0$ based on similarity of isotropic chemical shift to those found in the other pyrophosphate crystals which have a $\delta_{iso}$ between 1.3 to -6 ppm (Table S5 and Fig. S3). The peak position of $Q^1_0$ in the MATPASS and MAS deconvolution is at higher chemical shift (~ 4 ppm) because it is surrounded by more Li in its 2nd and 3rd coordination spheres as compared to the pyrophosphate crystals, a consequence of the global LiPON stoichiometry being closer to the orthophosphate $Li_3PO_4$. Such dependencies of $\delta_{iso}$ for $Q^n$ units based on the modifier content have been noted before in phosphate glasses and can vary the chemical shift upwards of 10 ppm.[34] The CSA parameters of the $Q^1_0$ site obtained from MATPASS are lower than that of the CSA values calculated for the two pyrophosphate crystals, $Li_4P_2O_7$, (Δδ ~ 130 ppm, η ~0.4) and the sign of Δδ is opposite of the crystals. This is partly due to the relative lower quantity of $Q^1_0$ sites found in LiPON in general and the large overlap of the $Q^0_0$ resonances complicating the extraction of the isolated $Q^1_0$ CSA parameters. Additionally, part of the $Q^1_0$ signal may not be completely refocused as noted above concerning why the $Q^0_1$ peak is not observed in the MATPASS experiment.

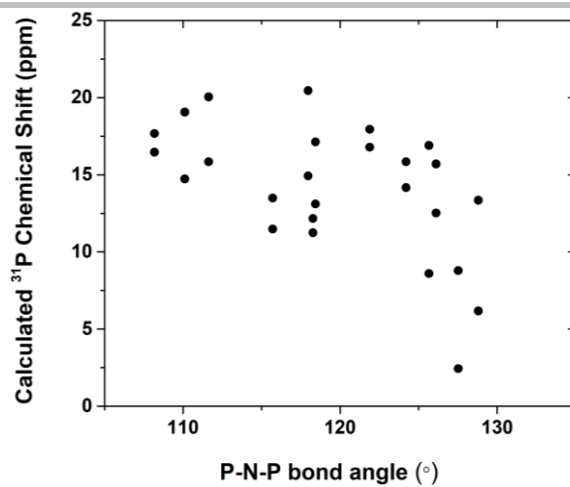

**Figure S7.** Bond angle variation of the $Q^1_1$ unit with $\delta_{iso}$. Displaying a clear trend towards lower chemical shift with higher bond angles. The variation is a source of broadening in the MAS spectrum and the cause of deviation from crystalline analogues.

To resolve the issues of incomplete refocusing of signal from the MATPASS experiment, we supplement the experimental CSA results with traditional side band analysis at various spinning speeds from single pulse experiments (Fig. S8). The free-standing film was used for the side band analysis. The width of the central band is relatively large (~20 ppm) requiring the spinning speed to be at least 5 kHz to separate the sidebands from overlapping with the central band. With increasing spinning speed, the intensity of the side bands diminishes rapidly, consistent with the low CSA values obtained from the MATPASS experiment. The CSA values for the two major peaks ($Q^0_0$ and $Q^1_1$) obtained from MATPASS are used to initially deconvolute the 5 kHz spectra with the remainder intensity in the sidebands from the $Q^0_1$ and $Q^1_0$ peaks. The CSA parameters for these two sites are then found to be: Δδ = -150 ppm and η = 0.3 and Δδ = -97.5 ppm and η= 0.3 for $Q^0_1$ and $Q^1_0$, respectively. The Δδ for $Q^1_0$ is larger than that obtained from MATPASS, indicating part of the signal was not refocused, however this larger Δδ is more consistent with the Δδ found in the pyrophosphate crystals as well as the $Q^1_0$ sites found in the AIMD structure. Despite the relatively low concentration of the $Q^0_1$ sites and the overlapping nature of the $^{31}$P LiPON spectrum, the large CSA for the $Q^0_1$ site is evident as it is the only remaining intensity contributing to the sidebands above 15 kHz. Combining the results from MATPASS and sideband analysis provides the complete CSA information of the four constituent structural units of LiPON. The average CSA values of the corresponding structural units from the AIMD generated structure are consistent with the experimental CSA values, providing essential justification for our assignments and deconvolution.

At 25 kHz, nearly all the CSA is averaged out, however, at this spinning speed it is clear there are additional contributions at higher chemical shifts, namely at 115 and 69 ppm. These peaks are not sidebands because they are independent of spinning speed (see 10, 15, and 25 kHz spectra) indicating they are isotropic chemical shifts. The identity of these resonances is unknown at this time and falls well beyond the isotropic chemical shift range observed for the oxynitride database presented here (the max $\delta_{iso}$ is found at ~48 ppm for $Li_7PN_4$), indicating these resonances are due to P in structural arrangements not considered within the crystal structures investigated. Rather the chemical shifts are close to those observed for phosphine (three coordinated P), where the 115 ppm shift is close to those of $PO_2N$ within organic molecules[35] and the 69 ppm shift is close to that of chemisorbed $PO_3$ found on zeolites.[36] Incorporating these sites into the deconvolution of the 25 kHz spectrum suggests these sites account for ~7% of the $^{31}$P signal, or roughly 7% of all P atoms within LiPON. These sites are tentatively assigned to phosphine units on the surface of LiPON and likely represent such a large fraction of the $^{31}$P signal because as a thin film the surface area to bulk volume ratio is relatively high. Whether these phosphine sites are an inherent surface structural feature of LiPON or a result of contamination on the exposed surface of the film requires further investigation. Surfaces of materials contain many structural defects and while those defects are less clearly defined in glasses, they can arise as under- or over-coordinated elements[37–39]; from this perspective these phosphine sites could be considered as surface defects. As such, they do not represent the 'bulk' structure of LiPON and are left out of the main text discussion as our AIMD model does not account for the structure of surfaces. However, observation of these potential surface sites is interesting as it implies MAS NMR can be a useful experimental technique to study solid-solid interfaces. The presence of these sites and their role in solid-solid interface reactions will be followed up in a future investigation.

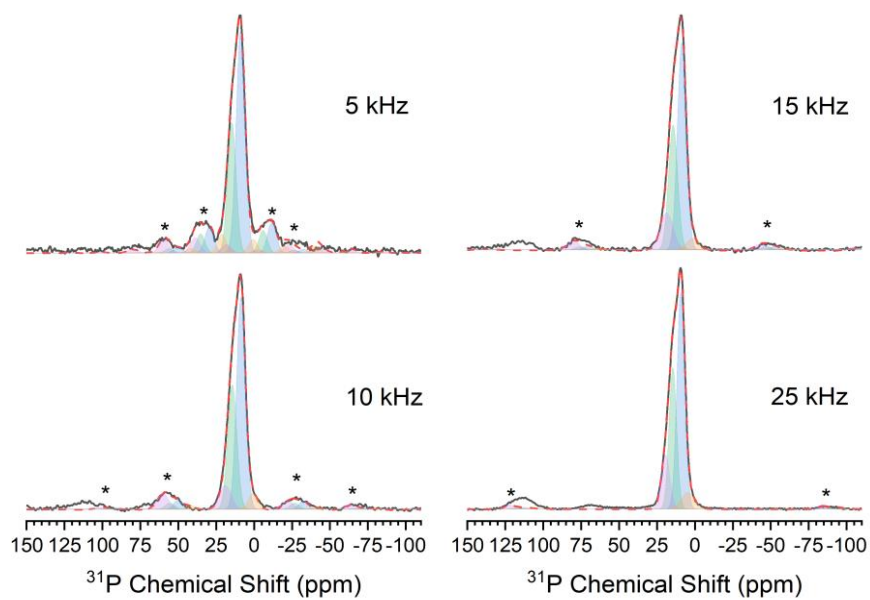

**Figure S8.** $^{31}$P MAS spectra at a variety of spinning speeds to analyze the sidebands for CSA parameters. Deconvolution parameters listed in Table S3. Asterisks denote spinning side bands. Note presence of an unidentified species at high chemical shift (~70 & 115 ppm) likely associated with surface defects.

### $^{31}$P Homonuclear Double Quantum Build-up Curves and Correlation Spectroscopy

Solid-state NMR is uniquely suited for studying the structure of materials, crystalline or amorphous, on both short and intermediate length scales. The CSA reflects the local symmetry of a phosphorous species on a short length scale and allows for different species to be identified based on their anisotropy. To supplement these assignments and gain insight into how these P species are distributed throughout the network, intermediate range effects like dipolar coupling need to be considered. Measuring dipolar coupling with NMR is an invaluable technique for studying the short and intermediate range ordering as it is directly related to the spatial separation between two nuclei. Investigating the internuclear distances between phosphate tetrahedra by $^{31}$P-$^{31}$P homonuclear double-quantum (DQ) NMR can reveal the ordering of the tetrahedra on intermediate length scales. While these are not measurements of bond distances, these spatial distances provide information analogous to that obtained from partial pair-distribution functions. These DQ measurements have been used to great effect to investigate the structure of phosphate glasses to determine the $Q^n$ connectivities for short and intermediate-range order, chain length distributions, and if the topological arrangement of the phosphate species are random or clustered.[40–48].

The information dipolar coupling provides can resolve some longstanding questions concerning the structure of LiPON. The literature exploring 'LiPON' materials suggest the RF sputtered orthophosphate based LiPON has extended chain structures through bridging N and O.[22,49–51] These claims typically rely on the observations within metaphosphate glasses which are already composed of extended chains. While these observations are helpful to understand the effect of nitridation, they cannot be applied to orthophosphate glasses like LiPON,[52] which are not expected to contain significant number of bridging O or N based on the starting composition. The only scenario in which a close to orthophosphate LiPON composition (the one that displays the enhanced stability against Li metal) can have extended chain structures is if the structure phase separates to form regions rich in Li and rich in P chains. However, the CSA analysis does not indicate any species like $Q^2_m$ that would lead to chain-like structures exist within LiPON, it is only composed of $Q^0_m$ and $Q^1_m$ species. The DQ NMR experiment can corroborate the findings from the CSA analysis and be used to determine if there is any clustering of the dimer units, which could lead to structures resembling extended chains on longer length scales.

The DQ build up curves and DQSQ correlation measurement were obtained with the BaBa-xy16 pulse sequence as it has good performance for refocusing relatively weak homonuclear dipolar coupling constants.[5] The pulse sequence records a spectrum with and without the double quantum filter, labelled DQ and MQ, respectively. Representative MQ and DQ spectra with an excitation time of 0.8 ms are shown in Fig. S9a, where the MQ spectra resembles the 1D MAS spectrum while the DQ spectrum shows much lower intensity and lacks the prominent peak at ~ 10 ppm. Connectivities with different dipolar coupling constants should display different rates of DQ build up, for example a $Q^0_m$ is generally further from neighboring P nuclei in comparison to a $Q^1_m$ or $Q^2_m$ thus it has a weaker homonuclear dipolar coupling constant and its build up will be slower. However, the peak resolution for the different $Q^n_m$ species are not clearly resolved due to broadening from chemical shift distribution. In order to observe whether there are any differences in the sites build up curves we integrate the chemical shift range expected for the four identified sites shown by the colored regions in Fig. S9a. The areas are used to plot the intensity functions (Fig. S9b) and normalized by the first MQ area. The MQ intensity shows a decay with increasing excitation time and the DQ intensity shows a gentle build up. The buildup curves are then normalized following the procedures outlined by Saalwächter, except tail subtraction was not performed as the long-time exponential tail of the MQ intensity is not reached within the excitation times collected. The normalized build up curves display typical behavior where there is a rapid rise in intensity before plateauing. However, the plateau is far below the DQ intensity of 0.5 that is expected in the long-time limit.[53] This is likely due to a dipolar truncation effect from the relatively strong heteronuclear couplings from $^7$Li and $^{14}$N that interfere with the weaker $^{31}$P-$^{31}$P homonuclear recoupling.[54,55] For example the estimated homonuclear $^{31}$P-$^{31}$P dipolar coupling constant (DCC) is on the order of 700 to 300 Hz for a distance of 3 to 4 Å while the heteronuclear $^{31}$P-$^7$Li DCC is on the order of ~1200 Hz for a bond distance of 2.5 Å and $^{31}$P-$^{14}$N DCC of ~800 Hz for a bond distance of 1.63 Å. The dipolar truncation is expected to impact the $Q^0_1$ and $Q^1_1$ more than the others because they are directly bonded to N. However, the large amount of Li relative to P will attenuate the recoupling of all $^{31}$P-$^{31}$P dipolar coupling signals and underestimates their DCC. To minimize the effect of these other coherences, the initial rise of the normalized DQ intensity can be fit (Fig. S9c) as it is dominated by DQ coherences.[5] Fitting the initial rise will provide an estimate of the apparent dipolar coupling, $D_{app}$, which encompasses the dipolar contributions from all spins in this multi-spin system. The initial rise, which only includes the first three data points can be fit with a second moment approximation:

$$I_{nDQ} \approx \frac{1}{2}\left(1 - e^{-2\left(\frac{6}{5\pi^2}D^2\tau_{DQ}^2\right)}\right)$$

Where $\tau_{DQ}$ is the DQ evolution time and D is the apparent dipolar coupling constant.

The fitting results reveal the $D_{app}$ values obtained for the four areas do not vary much and the two higher and lower chemical shift areas have approximately the same $D_{app}$ of ~270 Hz and ~210 Hz, respectively. These $D_{app}$ value for the lower chemical shift areas (12-2 ppm) corresponds to $^{31}$P-$^{31}$P distance of 4.48 Å. This distance is close to that expected for $Li_3PO_4$ (P-P distance of 4.1 Å), where all the P tetrahedra are isolated from one another. The $D_{app}$ value for the higher chemical shift areas (22-12 ppm) provides a slightly smaller P-P distance of 4.12 Å. This region is associated with P bonded to N in the form of $Q^0_1$ and $Q^1_1$, so the effect of dipolar truncation from $^{14}$N and $^7$Li are especially significant for this region and underestimate the true $D_{app}$ value. An estimate of the $^{31}$P-$^{31}$P $D_{app}$ for a $Q^1_1$ dimer can be obtained from the P-P distance of 2.9 Å in the $Li_5P_2O_6N$ crystal, providing 775 Hz. This $D_{app}$ value is slightly less than that expected of $^{31}$P-$^{14}$N $D_{app}$, so the $^{31}$P-$^{31}$P signal was attenuated by the latter. These results suggest that all P are well separated from one another and that the dimeric units do not represent a dominant fraction of the structural composition of LiPON. Otherwise their signal would not be attenuated as much. This result is largely consistent with the structural picture that MATPASS provides, where the predominate structural component in LiPON are $Q^0$ tetrahedra, followed by dimers with bridging N. Outside of the dipolar truncation effects suppressing the measurement of the N containing P tetrahedra, analysis of the interatomic P-P distances from the AIMD LiPON model largely agrees with the distances obtained from the build-up curves. Using a cut-off distance of 5.8 Å, the average global interatomic distance of neighboring P-P atoms is 4.87 Å ± 0.62Å. The results from the build-up curves indicates there is no considerable concentration of extended chain structures or tricluster units, as these would have much shorter interatomic P-P distances and they would be observable despite dipolar truncation effects if they were in high enough concentration. Rather the structure of LiPON is composed primarily of isolated tetrahedra and some dimeric units.

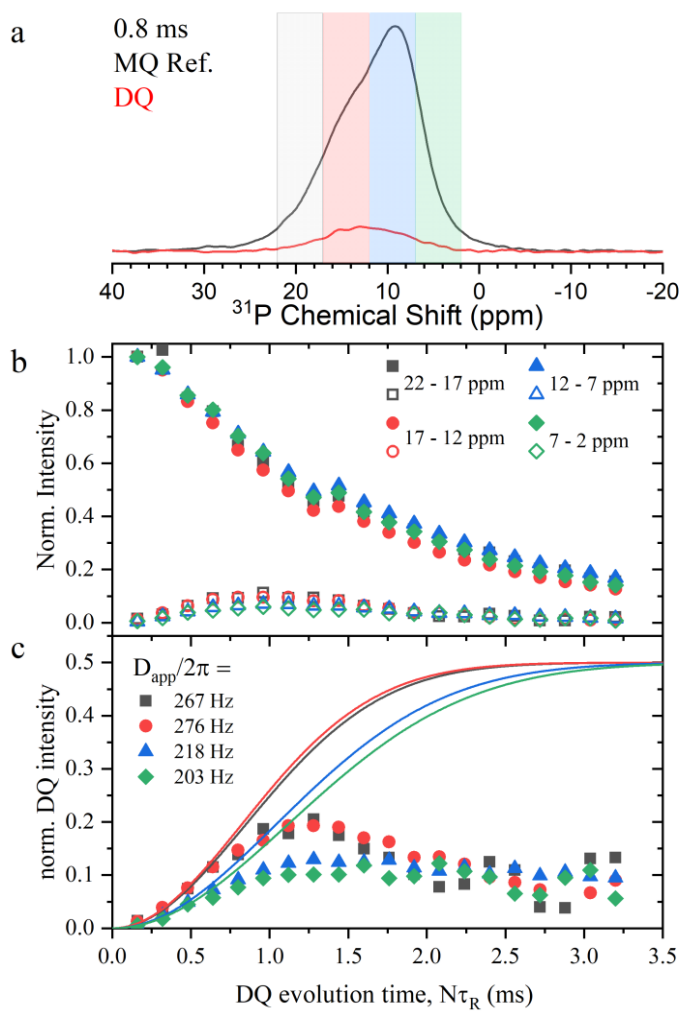

**Figure S9.** $^{31}$P-$^{31}$P DQ build up analysis. a) representative MQ and DQ spectra with the integrated regions marked by the colored areas, b) the MQ and DQ intensity functions, c) the normalized DQ intensity function with corresponding fit of the initial rise.

Further insight into the connectivity of these structural units can be obtained by double quantum single quantum (DQSQ) correlation measurements. The DQSQ correlation measurement can reveal which structural units are spatially close to others and more importantly if there is any clustering of any of the N containing Q units, as suggested by Sicolo et al.[50] The DQSQ contour plot reveals which structural units are in spatial proximity mediated through a dipolar interaction. Only sites with relatively strong dipolar couplings are shown in the SQ and DQ projections and weak dipolar couplings are not refocused.[56] If there is dipolar coupling between the P tetrahedra, a DQ peak will be present in the contour and in the projection of the DQ dimension corresponding to their sum frequency of their chemical shifts ($\omega_a + \omega_b$).[42] If the chemical shifts are the same, as in the same $Q^n$ unit is near another $Q^n$ unit of the same type, there will be a peak at double their chemical shift in the DQ projection ($2\omega_a$). These peaks are considered autocorrelated and fall on the DQ diagonal shown on the contour. Any correlations with dissimilar $Q^n$ units will display a cross-correlation peak off the diagonal at the sum frequency of the two chemical shifts ($\omega_a + \omega_b$) along the DQ projection. Analysis of these correlations can reveal which $Q^n$ units are connected to others. The DQSQ contour plot in Fig. S10 shows broad intensity smeared along the DQ diagonal, indicating that all P environments are auto-correlated with themselves. Additionally, there is broad intensity off the DQ diagonal, which is complicated by the chemical shift broadening from structural disorder. However, we can take the isotropic chemical shifts determined from the deconvolution of the MAS experiment and estimate the location of the auto correlations and cross correlations to determine if they are present. By carrying out this exercise, we see there is intensity for all the considered cross-correlations, indicating everything is homogeneously correlated and the P structural units are homogeneously distributed throughout the network. However, there are two exceptions, there is a lack of autocorrelation intensity for the $Q^0_1$ and $Q^1_0$ units, while they do display cross correlation intensity. This is because the $Q^0_1$ and $Q^1_0$ units are in much lower abundance and randomly distributed throughout the glass network, thus they have a very weak dipolar interaction with one another. This lack of intensity of the $Q^0_1$ unit is more apparent if the projection of the SQ dimension is compared to the MAS spectrum (Fig. S11). If all dipolar interactions are recoupled, then the two spectra will be identical. However, as the difference between the two spectra reveal, there is diminished intensity for the SQ projection above 18 ppm while there is non-insignificant intensity in the 1D MAS spectra extending upwards of 24 ppm. This lack of intensity provides additional support that the peak at 19 ppm is correctly assigned to $Q^0_1$. If this site were associated with $Q^N$ species greater than $Q^0$, it would have a stronger dipolar coupling with the neighboring P tetrahedra and would have an autocorrelation peak.

It is worth qualitatively comparing these results to P-P interatomic distances obtained from the AIMD LiPON model. As DQ projection closely mirrors the shape of the SQ projection, this implies all P sites are homogeneously mixed, so they are correlated with one another. This observation along with the results from the build-up curves (Fig. S9), indicates the average interatomic P-P distance between any like $Q^n_m$ pair is similar to the interatomic P-P distance of unlike pairs. The AIMD LiPON model reveals the average interatomic distances are not that different for like pairs of the four main structural units: $Q^0_0$-$Q^0_0$ is 4.89 Å ± 0.49Å, $Q^1_1$-$Q^1_1$ is 4.40 Å ± 1.05Å, $Q^0_1$-$Q^0_1$ is 5.15 Å ± 0.16Å, and $Q^1_0$-$Q^1_0$ is 2.95 Å ± 0.03Å. Notably, the P-P distance for the $Q^1_0$-$Q^1_0$ represents the intraatomic distance between the two P in a dimer, and interatomic distance is not measured because the $Q^1_0$ units are separated from one another due to their low concentration. In contrast to this, the $Q^1_1$ units are in much higher concentration and thus the average P-P distance includes both intra- and inter-atomic distances. It is important to note the interatomic distance of the $Q^1_1$ units is close to that of the $Q^0_0$ units, indicating most structural units are well separated from one another. These P-P values are in good agreement with those derived from partial pair distribution functions from neutron scattering.[15] Overall, the results of the AIMD LiPON agree with the results from the DQ based experiments and reveal there is no clustering of any of the $Q^n_m$ units within LiPON. Rather, the predominantly $Q^0_0$ and $Q^1_1$ units are well separated from one another and randomly distributed through the glass network.

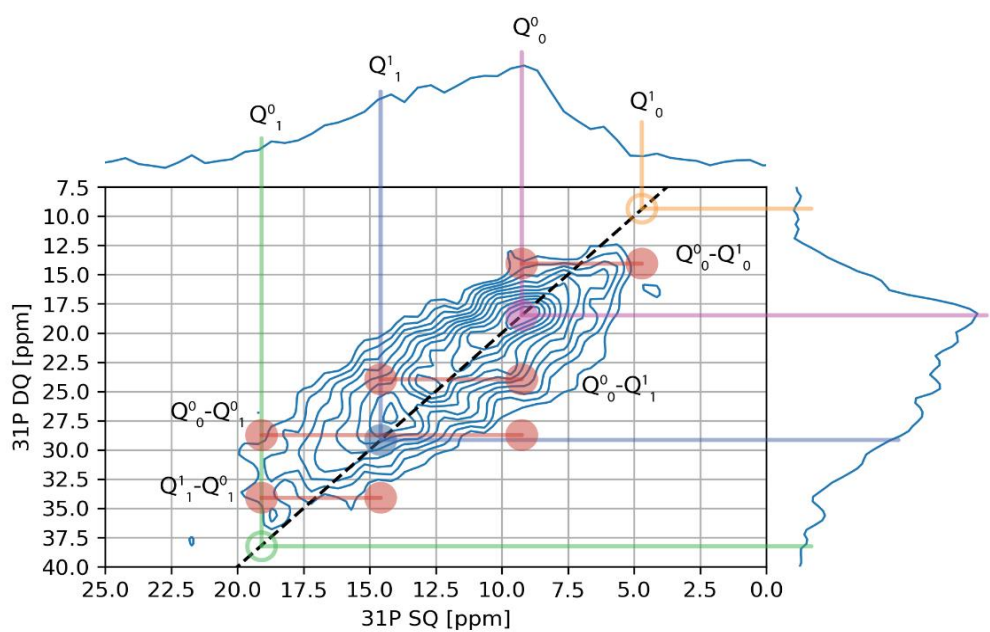

**Figure S10.** $^{31}$P DQSQ correlation contour plot. The DQ diagonal is shown by the dashed line with coloration corresponding to those in Fig. 1 to denote the chemical shifts in the SQ and DQ projections. Potential cross-correlation peaks are shown in red.

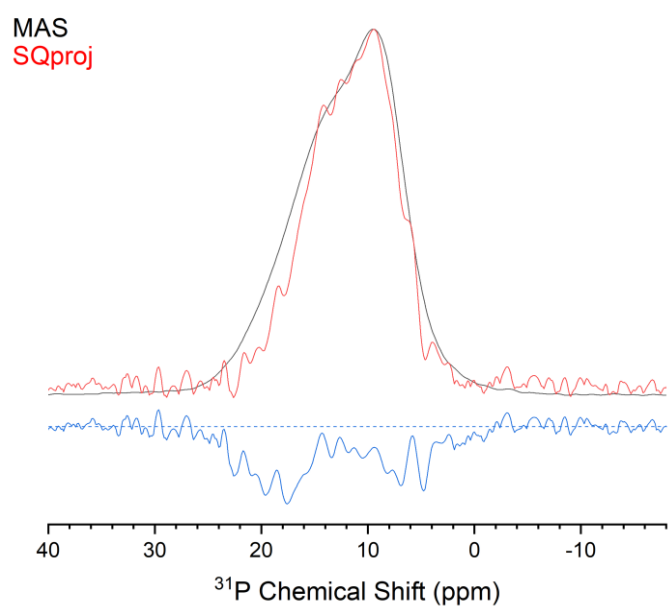

**Figure S11.** Comparison between a the $^{31}$P MAS line shape (Fig. 1) to the SQ projection, the difference of the two spectra shown below.

**Computational exploration of lithium oxynitride phosphates**

A database of lithium oxynitride phosphates were evaluated to observe the impact of local structure on the electronic shielding (Fig. S12). Experimental $^{31}$P chemical shifts reported in literature for the corresponding structures were used to correlate the calculated isotropic chemical shielding $\sigma_{calc}$ to experimental isotropic chemical shift $\delta_{exp}$. All reported experimental $^{31}$P chemical shifts exhibit a linear trend with their calculated shielding, barring the shifts reported by the works of Bertschler, et al., who investigated many of the lithium phosphorous nitride compounds.[57–59] These experimental shifts are systematically offset by ~18 ppm from the rest of the values in the database and they do not indicate how they referenced their $^{31}$P spectra and are omitted from consideration. Though, by omitting the compounds from Bertschler et al., the fit is not impacted greatly. A good linear correlation ($R^2$ = 0.976) between $\sigma_{calc}$ to $\delta_{exp}$ is achieved with a slope of 0.756 and a reference shielding value of 221.3. The compounds used and their computed parameters are provided in Tables S4-S7. In comparison to other studies calculating $^{31}$P chemical shifts using DFT methods, our slope and reference shielding values are in good agreement within ranges found from other studies exploring entirely different sets of inorganic phosphate compounds, indicating the calculation parameters used here are accurate at describing the underlying structure.[60,61] With an accurate linear correlation between $\sigma_{calc}$ and shift $\delta_{exp}$, we can convert the calculated chemical shieldings to chemical shifts, $\delta_{calc}$, for investigating structural trends influencing isotropic chemical shifts and CSA. This is especially useful for the phosphate compounds that have yet to be measured with $^{31}$P NMR.

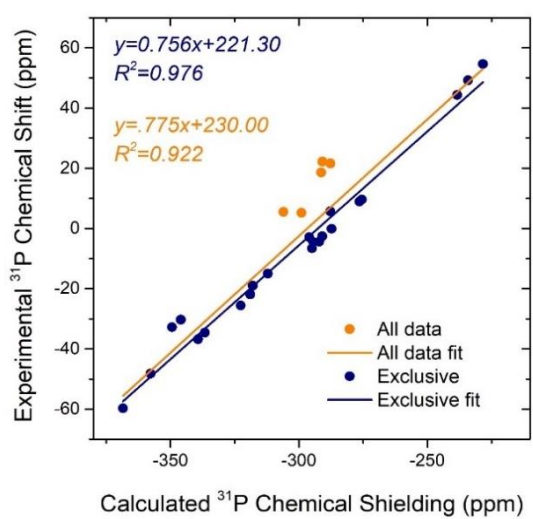

**Figure S12.** Comparison of calculated chemical shielding and experimental chemical shifts of lithiated phosphates and phosphate nitrides with (orange) and without (blue) data from Bertschler et al.

**Table S4.** Computational and structural details of the phosphate database.

| Structure | Supercell Info | | | | | | Source | Ref. | k-points | | |
|---|---|---|---|---|---|---|---|---|---|---|---|
| | a | b | c | α | β | γ | | | | | |
| AlPO$_4$ | 5.055 | 5.055 | 11.162 | 90 | 90 | 120 | Mat. Proj. | [62] | 5 | 2 | 2 |
| Ca(H$_2$PO$_4$)$_2$*(H$_2$O) | 5.643 | 6.556 | 12.014 | 98.39 | 83.53 | 118.15 | ICSD | [62] | 5 | 3 | 5 |
| P$_3$N$_5$ | 8.186 | 5.895 | 9.241 | 90 | 115.70 | 90 | | [63] | 3 | 4 | 2 |
| Cristobalite PON | 4.614 | 4.614 | 6.999 | 90 | 90 | 90 | ICSD | | 5 | 5 | 5 |
| δ-PON | 4.614 | 4.614 | 6.999 | 90 | 90 | 90 | ICSD | | 5 | 5 | 5 |
| P$_4$N$_6$O | 6.163 | 6.878 | 6.907 | 90 | 90 | 90 | Mat. Proj. | | 4 | 3 | 3 |
| LiPO$_3$ | 5.411 | 17.304 | 9.004 | 90 | 116.99 | 90 | | [64] | 4 | 4 | 3 |
| LiPN$_2$ | 4.571 | 4.571 | 7.290 | 90 | 90 | 90 | Mat. Proj. | | 5 | 5 | 3 |
| Li$_2$PO$_2$N | 9.156 | 5.461 | 4.749 | 90 | 90 | 90 | | [65] | 2 | 4 | 5 |
| Li$_3$P | 8.472 | 8.472 | 15.141 | 90 | 90 | 120 | | [58] | 4 | 4 | 2 |
| β-Li$_3$PO$_4$ | 12.352 | 10.593 | 9.845 | 90 | 90 | 90 | Mat. Proj. | | 2 | 2 | 2 |
| γ-Li$_3$PO$_4$ | 10.615 | 6.177 | 5.001 | 90 | 90 | 90 | | | 2 | 4 | 5 |
| Li$_4$P$_2$O$_7$ - pseudo-monoclinic | 5.260 | 14.118 | 9.571 | 90 | 123.40 | 90 | Mat. Proj. | [66] | 5 | 1 | 2 |
| Li$_4$P$_2$O$_7$ - triclinic | 5.261 | 7.215 | 8.662 | 77.33 | 89.98 | 68.88 | Mat. Proj. | [66] | 4 | 4 | 2 |
| Li$_4$PN$_3$ | 9.660 | 11.839 | 4.867 | 90 | 90 | 90 | | [58] | 5 | 5 | 3 |
| Li$_5$P$_2$N$_5$ | 14.770 | 17.850 | 4.860 | 90 | 93.11 | 90 | | [57] | 2 | 2 | 5 |
| Li$_5$P$_2$O$_6$N | 8.460 | 7.200 | 4.860 | 109.9 | 90.3 | 100.2 | Ref. [67] | [67] | 3 | 3 | 5 |
| Li$_7$PN$_4$ | 9.402 | 9.402 | 9.402 | 90 | 90 | 90 | Mat. Proj. | | 2 | 2 | 2 |
| Li$_{10}$P4N$_{10}$ | 12.309 | 12.309 | 12.309 | 90 | 90 | 90 | ICSD | | 2 | 2 | 2 |
| Li$_{13}$P4N$_{10}$ | 12.309 | 12.309 | 12.309 | 90 | 90 | 90 | | | 2 | 2 | 2 |
| Li$_{14}$(PON$_3$)$_2$O | 5.699 | 5.699 | 8.072 | 90 | 90 | 120 | | [58] | 4 | 4 | 3 |
| Li$_{18}$P$_6$N$_{16}$ | 5.426 | 7.535 | 9.858 | 108.48 | 99.29 | 105.00 | ICSD | [68] | 4 | 4 | 2 |
| Si$_3$N$_4$ | 7.661 | 7.661 | 2.925 | 90 | 90 | 120 | Mat. Proj. | | 3 | 3 | 9 |

**Table S5.** Computed $^{31}$P chemical shielding and experimental $^{31}$P chemical shifts, with $Q^n$ and $Q_m$.

| Structure | Calculated $^{31}$P Chemical Shielding | | | | Experimental $^{31}$P Chemical Shift | $Q^n$ | $Q_m$ | Ref. |
|---|---|---|---|---|---|---|---|---|
| | $\sigma_{iso}$ (ppm) | $\Delta\delta$ (ppm) | $\eta$ | $\delta_{iso}$, fit (ppm) | $\delta_{iso}$ (ppm) | | | |
| AlPO$_4$ | -322.653 | 4.391 | 0.834 | -22.607 | -25.60 | | | [60] |
| Ca(H$_2$PO$_4$)2*(H$_2$O) | -294.368 | 92.597 | 0.836 | -1.247 | -4.60 | | | [60] |
| | -287.266 | 72.400 | 0.936 | 4.117 | -0.10 | | | |
| P$_3$N$_5$ | -368.455 | -197.017 | 0.109 | -57.197 | -56.2, -59.7, -64.8 | 3 | 4 | [69] |
| | -357.723 | 85.590 | 0.251 | -49.093 | -46, -48.2 | 4 | 4 | |
| Cristobalite PON | -345.980 | -132.488 | 0.384 | -40.224 | -30.3 | 4 | 2 | [70] |
| δ-PON | -357.820 | -9.982 | 0.000 | -49.166 | | 4 | 4 | [70] |
| | -360.229 | -198.943 | 0.640 | -50.985 | | 4 | 2 | |
| | -349.380 | 0.462 | 0.000 | -42.791 | -32.8 | 4 | 0 | |
| P$_4$N$_6$O | -351.609 | -205.716 | 0.062 | -44.475 | | 4 | 4 | |
| | -355.721 | -90.209 | 0.433 | -47.580 | | 4 | 3 | |
| LiPO$_3$ - Rings | -312.104 | -191.672 | 0.440 | -14.641 | | 2 | 0 | [71] |
| | -318.954 | -218.293 | 0.423 | -19.814 | | 2 | 0 | |
| LiPO$_3$ - Chains | -317.955 | 241.039 | 0.551 | -19.059 | -15, -19, -21.9, -25.1 | 2 | 0 | [71] |
| LiPN$_2$ | -287.756 | -27.045 | 0.000 | 3.747 | 5.68 | 4 | 4 | [72] |
| Li$_2$PO$_2$N | -272.922 | -52.016 | 0.162 | 14.949 | | 2 | 2 | |
| Li$_3$P | -494.399 | -241.318 | 0.000 | -152.310 | -278.00 | | | [73] |
| β-Li$_3$PO$_4$ | -275.477 | -1.958 | 0.612 | 13.020 | 9.60 | 0 | 0 | Reported here |
| γ-Li$_3$PO$_4$ | -276.468 | 3.836 | 0.847 | 12.271 | 8.9 | 0 | 0 | [74] |
| Li$_4$P$_2$O$_7$ - pseudo-monoclinic | -295.965 | 131.960 | 0.438 | -2.453 | -2.60 | 1 | 0 | [27] |
| | -290.931 | 130.157 | 0.342 | 1.349 | | 1 | 0 | |
| Li$_4$P$_2$O$_7$ - triclinic | -292.043 | 128.896 | 0.367 | 0.509 | -4.4 | 1 | 0 | [27] |
| | -294.885 | 133.414 | 0.423 | -1.637 | -6.6 | 1 | 0 | |
| Li$_4$PN$_3$ | -287.730 | -148.798 | 0.302 | 3.766 | 21.58 | 2 | 4 | [57] |
| Li$_5$P$_2$N$_5$ | -335.463 | -167.519 | 0.286 | -32.282 | 5.22 | 3 | 4 | [57] |
| | -302.997 | -130.373 | 0.293 | -7.763 | | 3 | 4 | |
| | -298.986 | -140.261 | 0.261 | -4.734 | | 3 | 4 | |
| Li$_5$P$_2$O$_6$N | -290.493 | 57.702 | 0.381 | 1.680 | | 1 | 1 | |
| | -289.037 | 46.676 | 0.373 | 2.779 | | 1 | 1 | |
| Li$_7$PN$_4$ | -228.363 | -20.824 | 0.000 | 48.600 | 54.60 | 0 | 4 | [72] |
| | -234.161 | 0.002 | 0.956 | 44.222 | 49.20 | 0 | 4 | |
| Li$_{10}$P$_4$N$_{10}$ | -296.137 | -119.787 | 0.141 | -2.583 | 12.80 | 3 | 4 | [75] |
| Li$_{12}$P$_3$N$_9$ | -290.440 | -147.326 | 0.428 | 1.720 | 22.76, 15.08 | 2 | 4 | [58] |
| Li$_{13}$P$_4$N$_{10}$ | -302.562 | -164.540 | 0.000 | -7.435 | | 3 | 4 | |
| Li$_{14}$(PON$_3$)$_2$O | -238.399 | 99.479 | 0.000 | 41.021 | 44.30 | 0 | 3 | [76] |
| Li$_{18}$P$_6$N$_{16}$ | -306.039 | -146.626 | 0.355 | -10.060 | 5.5 | 3 | 4 | [59] |
| | -291.315 | -138.449 | 0.471 | 1.059 | 18.6 | 2 | 4 | |
| | -290.832 | -144.717 | 0.165 | 1.423 | 22.2 | 3 | 4 | |

Table S6. Computed [15]N chemical shielding and experimental [15]N chemical shifts.

| Structure | Calculated [15]N Chemical Shielding | | | | Experimental [15]N Chemical Shift | N Coordination | Reference |
|---|---|---|---|---|---|---|---|
| | $\sigma_{iso}$ (ppm) | $\Delta\delta$ (ppm) | $\eta$ | $\delta_{iso}$, fit | $\delta_{iso}$ (ppm) | | |
| a-Glycene | -189.014 | -12.743 | 0.401 | 34.583 | 33.4 | | [77] |
| $P_3N_5$ | -130.539 | -46.519 | 0.138 | 85.269 | 79.30 | 2 | [78] |
| | -80.493 | 26.771 | 0.622 | 128.649 | 129.30 | 3 | |
| | -145.088 | -86.104 | 0.221 | 72.657 | 79.30 | 2 | |
| Cristobalite PON | -107.952 | -13.741 | 0.686 | 104.847 | | 2 | |
| δ-PON | -120.732 | -47.810 | 0.735 | 93.770 | | 2 | |
| | -127.738 | -32.565 | 0.344 | 87.697 | | 2 | |
| | -125.107 | -20.520 | 0.423 | 89.977 | | 2 | |
| $P_4N_6O$ | -133.233 | -22.957 | 0.905 | 82.934 | | 2 | |
| | -92.384 | -41.172 | 0.700 | 118.342 | | 3 | |
| $LiPN_2$ | -147.755 | -33.163 | 0.629 | 70.346 | | 2 | |
| $Li_2PO_2N$ | -162.320 | -47.804 | 0.029 | 57.721 | | 2 | |
| $Li_4PN_3$ | -119.346 | 27.249 | 0.889 | 94.971 | | 1 | |
| | -75.841 | -40.384 | 0.734 | 132.681 | | 2 | |
| $Li_5P_2N_5$ | -126.594 | -34.831 | 0.647 | 88.688 | | 1 | |
| | -101.282 | -42.875 | 0.560 | 110.628 | | 2 | |
| $Li_5P_2O_6N$ | -131.942 | -33.589 | 0.263 | 84.053 | | 2 | |
| $Li_7PN_4$ | -152.597 | -18.955 | 0.209 | 66.149 | | 1 | |
| | -147.552 | -9.583 | 0.000 | 70.522 | | 1 | |
| $Li_{10}P_4N_{10}$ | -124.135 | 10.252 | 0.823 | 90.820 | | 2 | |
| | -105.602 | 44.375 | 0.729 | 106.884 | | 1 | |
| $Li_{12}P_3N_9$ | -107.598 | 40.790 | 0.855 | 105.154 | | 1 | |
| | -103.882 | 26.355 | 0.621 | 108.375 | | 2 | |
| $Li_{13}P_4N_{10}$ | -113.65 | 27.795 | 0.074 | 99.910 | | 2 | |
| | -178.91 | -50.680 | 0.000 | 43.342 | | 1 | |
| $Li_{14}(PON_3)_2O$ | -146.368 | 39.507 | 0.612 | 71.548 | | 1 | |
| $Li_{18}P_6N_{16}$ | -122.326 | 36.227 | 0.841 | 92.388 | | 1 | |
| | -104.858 | -55.034 | 0.499 | 107.529 | | 2 | |
| $Si_3N_4$ | -123.634 | 37.946 | 1.000 | 91.254 | 90.90 | 3 | [79] |
| | -144.078 | -25.885 | 0.984 | 73.533 | 73.70 | 2 | |

Table S7. Computed [6/7]Li chemical shielding and experimental [6/7]Li chemical shifts.

| Structure | Calculated [7]Li Chemical Shielding Data | | | Experimental [6,7]Li Chemical Shift Data (ppm) | Reference |
|---|---|---|---|---|---|
| | $\sigma_{iso}$ (ppm) | $\Delta\delta$ (ppm) | $\eta$ | | |
| $LiPO_3$ | -58.811 | 4.101 | 0.402 | | |
| | -58.266 | 8.305 | 0.378 | | |
| | -58.446 | -4.440 | 0.885 | | |
| $LiPN_2$ | -57.392 | 3.266 | 0.000 | [6]Li: 1.64/[7]Li: 1.66 | [57] |
| $Li_2PO_2N$ | -57.449 | 2.216 | 0.937 | | |
| $Li_3P$ | -52.634 | -8.485 | 0.001 | 0.40 | [73] |
| | -49.849 | 8.569 | 0.001 | 4.70 | |
| $\beta$-$Li_3PO_4$ | -57.157 | -2.428 | 0.394 | 0.137 | Reported here |
| | -57.272 | 3.533 | 0.824 | | |
| $\gamma$-$Li_3PO_4$ | -57.195 | 3.306 | 0.686 | | |
| | -57.267 | 5.061 | 0.551 | | |
| $Li_4P_2O_7$ - pseudo-monoclinic | -57.753 | 2.848 | 0.923 | | |
| | -57.869 | -3.218 | 0.703 | | |
| | -57.683 | 9.920 | 0.559 | | |
| $Li_4P_2O_7$ - triclinic | -57.310 | 2.929 | 0.802 | | |
| | -57.264 | 9.871 | 0.616 | | |
| | -57.089 | -4.661 | 0.325 | | |
| | -57.406 | -2.921 | 0.775 | | |
| $Li_4PN_3$ | -86.636 | -6.051 | 0.849 | 2.59 | [57] |
| $Li_5P_2N_5$ | -87.808 | -2.012 | 0.287 | [6]Li: 5.43, 1.87 [7]Li: 4.9, 1.7 | [57] |
| $Li_5P_2O_6N$ | -88.344 | -5.360 | 0.825 | | |
| $Li_7PN_4$ | -54.614 | -4.489 | 0.746 | 3.33 | [72] |
| $Li_{10}P_4N_{10}$ | -86.407 | 9.239 | 0.642 | 2.3, 1.8, -0.5 | [75] |
| $Li_{12}P_3N_9$ | -86.855 | -10.863 | 0.267 | | |
| | -87.106 | -9.115 | 0.843 | | |
| | -86.860 | 5.890 | 0.843 | | |
| | -86.546 | 8.040 | 0.412 | | |
| $Li_{13}P_4N_{10}$ | -89.561 | -5.730 | 0.000 | | |
| | -81.362 | 0.002 | 0.493 | | |
| | -86.092 | 16.079 | 0.275 | | |
| | -78.899 | -18.380 | 0.000 | | |
| $Li_{14}(PON_3)_2O$ | -55.360 | 7.413 | 0.553 | 2.34 | [76] |
| $Li_{18}P_6N_{16}$ | -87.188 | 5.914 | 0.891 | 1.60 | [59] |

# SUPPORTING INFORMATION

**$^{15}$N results from DFT**

During deposition of the LiPON films on fused quartz rods, $^{15}$N enriched gas (98%) was used as the N source. This enrichment was done to measure the $^{15}$N signal of LiPON to supplement the $^{31}$P data as an indicator of the local structure. However, we were unsuccessful in observing any meaningful $^{15}$N signal that rose above the noise likely because $^{15}$N is a low gamma nuclei making it challenging to observe in the first place but also LiPON as a thin film is sample limited and with 20% enrichment, $^{15}$N makes up less than 1 at% of LiPON. While our attempts to measure a $^{15}$N NMR signal were unsuccessful, there have been previous measurements of $^{15}$N in nitrided glasses, including LiPON synthesized using ion beam assisted deposition (IBAD).[80,81] Bunker et al. studied the effect of nitridation on sodium metaphosphate glasses and found two $^{15}$N peaks whose relative intensities remain invariant with N content however their chemical shifts slightly increase with increasing N content. They assigned the peak at 112 ppm to the triple coordinated N, $N_t$, in a -N< environment and 77 ppm to the double coordinated or bridging N, $N_d$, a bridging anion between P tetrahedra P-N=P. Stallworth et al. performed $^{15}$N NMR on LiPON synthesized using IBAD and obtained a broad peak at 85.3 ppm which they assign to P-N=P linkages and a sharp peak at 315.3 ppm associated with entrapped $N_2$ gas molecules. The GIPAW calculations also can be used to provide insight into the relationship of structures measured with $^{15}$N and can be used to evaluate previous findings in a new light. These can also be used in conjunction with the AIMD generated structure to directly estimate the $^{15}$N chemical shift of our LiPON structure. All discussions of $^{15}$N chemical shifts, including prior reports, have been converted to the 'unified scale' reported by Bertani et al.[77] While few reports of $^{15}$N chemical shifts exist for the relevant compounds in this database, we performed additional calculations on α-glycine to use as a reference for the unified scale and $Si_3N_4$ for consideration of another compound where $N_t$ is found.

It should be noted, there is a pervasive assumption within literature on oxynitride glasses that $N_t$ is a dominant structural feature followed by $N_d$, with little consideration of direct substitution on isolated $PO_4$ orthophosphate groups referred to as apical sites, $N_a$. This assumption is largely due to initial tentative assignments based on XPS results without any further validation, as discussed by Lacivita et al.[15] We will expand on this discussion with some observations from the GIPAW database calculations performed on N containing phosphate compounds. First, it is worth considering if there is any precedent for $N_t$ to form at all within the Li-P-O-N phase space. From our extensive literature search on relevant phases used to construct the database, we found only two phosphorous compounds, $P_3N_5$ and $P_4N_6O$, contain $N_t$. These are isostructural to one another with the only difference is an O substituted on a bridging site. The $N_t$ site only appears at the vertex of an edge-sharing structure and at the corner of a third P tetrahedra[82], in direct contrast to the $N_t$ site found in $Si_3N_4$ in which N is the vertex corner of three Si tetrahedra. This suggests that if $N_t$ were to form in an oxynitride phosphate compound it will be accompanied by edge sharing P tetrahedra. All compounds within our database consist exclusively of corner sharing tetrahedra, and there have not been any reports, as to the authors knowledge, of edge-sharing P tetrahedra in alkali phosphate glasses.[28,83]

However, another important consideration is these two $N_t$ containing compounds have no alkali for charge balancing and suggests the formation of $N_t$ may have a dependence on alkali content. While there is a lack of oxynitride compounds with variable Li content, there is a wide range for LiPN compounds ($LiPN_2$, $Li_4PN_3$, $Li_5P_2N_5$, $Li_{10}P_4N_{10}$, $Li_{18}P_6N_{16}$, $Li_7PN_4$). These compounds show a variety of topological structures based entirely on corner sharing tetrahedra ranging from a 3D network, sheets, chains, rings, and isolated tetrahedra that vary with the alkali content.[72] Interestingly, a structure based on dimers has not been reported, whereas dimeric structures have been observed for pyrophosphate compounds and for a theoretical pyrophosphate oxynitride. The variability of structural units lacking formation of edge sharing P suggests that $PN_4$ tetrahedra prefer to form corner sharing configurations so long as cations are available for charge compensation. As no edge sharing units are observed for LiPN or LiPO compounds there is no reason to expect edge-sharing P within the lithium oxynitride phosphate compounds. That said, this does not rule out the existence of edge-sharing P from forming within a glass network as metastable structural units otherwise not found in low pressure/temperature crystals have been found in glasses.[39]

Many prior studies have on oxynitride glasses have looked at metaphosphate compositions and discuss N incorporation in terms of substitution of O anion sites causing a net polymerization of the network. It has been assumed that because $N_t$ is observed within oxynitride silicate glasses with XPS it must also be present in oxynitride phosphate glasses. However, N incorporation into modifier rich compositions greater than metaphosphates could result in different structural units due to the excess of cations for charge compensation. For example, if N is substituted on a bridging O site in a metaphosphate glass ($2[PO_3]^{1-} + N \rightarrow [P_2O_5N]^{3-}$), the net charge changes from -2 to -3 indicating a charge imbalance. This can be resolved by including a third $PO_3$ unit and forming $N_t$ ($3[PO_3^{1-}] \rightarrow P_3O_8N^{3-}$), creating a charge balanced tricluster of P tetrahedra. Note that one oxygen vacancy is created for every 2 N substituted onto the network.[84] This gives some precedence for $N_t$ forming within metaphosphate glass compositions. However, the scenario is very different for an orthophosphate composition, that closest to LiPON thin films. Here N can only substitute on a non-bridging O site or the terminal O (π bond) site, although it does not matter which site N substitutes for the purposes of charge counting. If we perform a similar N substitution to create a tricluster ($3[PO_4]^{3-} + N \rightarrow [P_3O_9N]^{6-}$), this results in a net charge of the tetrahedral cluster going from -9 to -6 along with the creation of 2 O vacancies. This situation implies the formation of a tricluster, and implicitly $N_t$, will result in charge imbalances for an orthophosphate glass. Rather, introducing a second N atom and forming a dimer and substituting N on either the double bond or non-bridging oxygen site of the third P tetrahedra conserves the net charge ($3[PO_4]^{3-} + 2N \rightarrow [P_2O_6N]^{5-} + [PO_3N]^{4-}$) and forms only one O vacancy. Alternatively, the 2N could be on the same dimer and still maintain the net charge and creation of one O vacancy ($2[PO_4]^{3-} + 2N \rightarrow [P_2O_5N_2]^{6-}$). This scenario implies that substitution of N into an orthophosphate compound is much more likely to result in the formation of a dimeric unit or create non-bridging N along with dimers in a 1:1 ratio. The formation of dimers and $N_a$ is supported by the findings of Wang et al. where they observed both types of N incorporation in a nitrided LiPON composition close to $Li_3PO_4$.[85] The energetics of forming the P-N-P dimers in $Li_3PO_4$ have been found to be lower energy than forming $N_a$, indicating there may be a deviation from the 1:1 ratio suggested by the charge balancing exercises.[67]

From these observations we can reject the assumption that $N_t$ is expected to form in every lithium phosphate compound or glass and propose that it may be compositionally dependent. From $^{15}N$ NMR data, the $N_t$ and $N_d$ site in $P_3N_5$ have been measured with $^{15}N$ NMR, displaying an isotropic chemical shift of 129.3 ppm for $N_t$ and 79.3 ppm for $N_d$, indicating there is a chemical shift difference of about 50 ppm.[78] Each of the LiPN compounds from our GIPAW calculations (except for the end members $LiPN_2$ and $Li_7PN_4$, which are exclusively composed of $Q^4$ and $Q^0$, respectively) also have two $^{15}N$ chemical shifts, typically in the ranges of 88-105 and 107-133 ppm. The difference between these shifts decreases with increasing Li content from 37 ppm to 3 ppm, suggesting the N environment is becoming increasingly similar. Given the absence of any $N_t$, the two $^{15}N$ shifts found for the LiPN compounds can be clearly assigned to $N_a$ (88-105 ppm) and $N_d$ (107-133 ppm). The chemical shift of the bridging N appears to roughly decrease as the Li content increases and the chemical shift for non-bridging N does not have a clear compositional correlation with Li. The effect of $Q^n$ and N coordination on $^{15}N$ chemical shift is plotted in Fig. S13. There is no clear relationship on the $^{15}N$ chemical shift with $Q^n$, however there is a rough trend correlating decreased $^{15}N$ shift with decreased N coordination, although there is considerable variation. While we lack a systematic dataset to make any general trends of how $Q^n$ and/or $Q_m$ influences $^{15}N$ chemical shift of either $N_a$ or $N_d$, there are two notable oxynitride examples from the GIPAW database, $Li_2PO_2N$ and $Li_5P_2O_6N$, which are isostructural to a metaphosphate ($LiPO_3$) and pyrophosphate ($Li_4P_2O_7$) except for N occupying the bridging site rather than O. Both compounds only have one $^{15}N$ chemical shift, 57 ppm for $Li_2PO_2N$ and 84 ppm for $Li_5P_2O_6N$, and both with $N_d$. First, $Li_2PO_2N$ can be compared to $Li_4PN_3$ as the local structure is the same, comprised of tetrahedral chains connected by P-N=P linkages. The only differences are the non-bridging sites are all O in $Li_2PO_2N$ and all N in $Li_4PN_3$ and the additional Li for charge compensation. The chemical shift for $Li_2PO_2N$ is dramatically lower in comparison to the bridging site in $Li_4PN_3$ (132 ppm), suggesting that increasing the number of O on the P tetrahedra lowers the $^{15}N$ shift of the bridging N. The $^{15}N$ chemical shift for $Li_5P_2O_6N$ is interesting to consider because it represents the chemical shift of a bridging N in a P-P dimer, which is expected to be the major structural unit of N in LiPON (See main text, and Lacitiva et al.[15]). In comparison to $Li_2PO_2N$ it has a higher shift, although structurally it is different as it bonded in a dimer rather than a chain ($Q^1$ vs $Q^2$) and has one more oxygen per tetrahedra. It is clear from this comparison, even for structures with the same coordination of N, $N_d$, the $^{15}N$ shift has a wide distribution (57 to 132 ppm) that varies with the particular structure, Li content, topology, orientation of the double bond on P, and O per P tetrahedra. We need careful systematic studies using a combined approach of experimental NMR, GIPAW calculations, and other complimentary methods to definitively verify the relationship of the local environment to $^{15}N$ chemical shift.

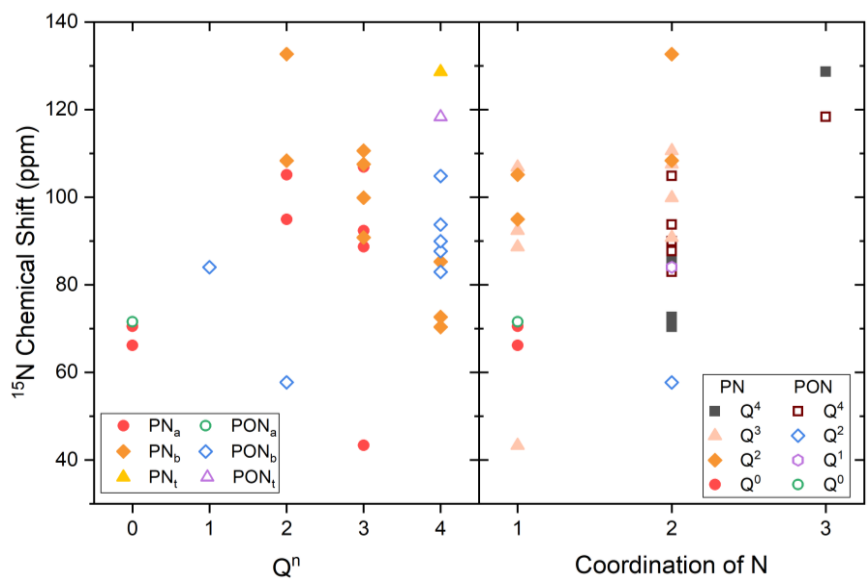

**Figure S13.** Calculated $^{15}N$ chemical shift plotted as a function of $Q^n$ (left) and N coordination (right).

Despite the complexity of the $^{15}N$ chemical shifts, with the current dataset and observations we can offer an alternative interpretation of the $^{15}N$ data obtained by Bunker et al. for their sodium metaphosphate glasses. Rather than $N_t$ and $N_d$, the two peaks can be assigned to $N_d$ in a $Q^2$ chain at 112 ppm and a non-bridging N, $N_a$, at 77 ppm. As a general consideration, the $^{15}N$ chemical shift increases with N coordination and the range for observed $N_t$ chemical shifts is above the 112 ppm peak, indicating it could be assigned to a double bridging site rather than $N_t$. It should be noted however that the $N_t$ chemical shift found in $P_4N_6O$ is 118 ppm and the $N_d$ chemical shift is 83 ppm, which are very close to the chemical shifts obtained by Bunker et al.[80] The assignments of the original $N_t$ and $N_d$ or $N_d$ and $N_a$ are equally valid from a $^{15}N$ NMR perspective, though requires further validation by carefully analyzing whether edge-sharing P exists in these nitride glasses. If edge-sharing is present in P, there should be a corresponding broadening or separate peak in the $^{31}P$ chemical shift. However, no new peak is observed in the Bunker et al. data, suggesting N randomly substitutes onto already existing O sites, (bridging, non-bridging, and terminal) rather than creating a new three coordinated bridging vertex.

# SUPPORTING INFORMATION

In contrast to the $^{15}$N NMR results observed in the sodium metaphosphate glasses, the $^{15}$N spectrum for the IBAD LiPON only has one peak at 85.3 ppm associated with N in its structure. This peak is in good agreement with the calculated shift for the dimeric P-N=P linkage found in the Li$_5$P$_2$O$_6$N compound ($\delta_{iso}$= 84 ppm), indicating that the N in IBAD LiPON only forms P-N=P linkages rather than forming both N$_a$ and N$_d$. However, the corresponding $^{31}$P spectra also lacks intensity above 18 ppm, where the $Q^0_1$ $^{31}$P chemical shift is found (see main text). As suggested by modeling studies on Li$_3$PO$_4$, N is more stable as a bridging dimer than at an apical site, so the higher energies from the IBAD synthesis route could have destroyed the apical sites, whereas these sites are retained during RF sputtering. This implies both N$_a$ and N$_d$ $^{15}$N chemical shifts should be present within the RF sputtered LiPON. We can use the calculated shifts from the AIMD LiPON structure to estimate the N environments. Just as the $^{31}$P NMR analysis suggests, N is found in both N$_a$ and N$_d$ in the AIMD LiPON structure. The resulting $^{15}$N chemical shifts however are somewhat similar and have a very broad range, with the average chemical shift of 60.9 ± 7.1 ppm for N$_a$ and 66.9 ± 6.99 ppm for N$_d$. These values agree with the observations made in the GIPAW database with there being a lower shift with lower N coordination and a decreasing difference in the isotropic shift with a high Li content. Given the broad distribution of chemical shifts and their similarity, it is unlikely these two sites could be resolved in an experimental spectrum. However, their CSA parameters display notable differences with the average CSA parameters of Δδ= -24 ppm and η=0.95 for N$_a$ and Δδ= -50 ppm and η= 0.36 for N$_d$, and could be used to distinguish that both sites are present by careful analysis of the CSA.

# SUPPORTING INFORMATION